\documentclass[a4paper]{article}

\usepackage[english]{babel}
\usepackage[utf8]{inputenc}
\usepackage[scaled]{helvet}
\usepackage[T1]{fontenc}

\usepackage[a4paper,top=3cm,bottom=2cm,left=3cm,right=3cm,marginparwidth=2cm]{geometry}

\usepackage{amsmath} % math package
\usepackage{amsfonts} % solve "Undefined control sequence" error of mathbb
\usepackage{graphicx}
\usepackage{tabularx}
\usepackage{longtable}
\usepackage{algorithm}
\usepackage{algorithmic}
\usepackage{booktabs}
\usepackage[section]{placeins} %% Keep order for tables, images
\usepackage{natbib}
\usepackage[colorlinks=true, allcolors=blue]{hyperref}
\setcitestyle{authoryear,open={(},close={)}}
\date{}

\title{Expert Aggregation for Financial Forecasting}
\author{Carl Remlinger\footnote{EDF lab; FiME Laboratory} \footnote{Universit\'e Gustave Eiffel}, Clémence Alasseur, Marie Brière\footnote{Amundi Asset Management; Paris Dauphine PSL University; Université Libre de Bruxelles}, Joseph Mikael}

\begin{document}

%%%%%%%%%%%%%%%%%%%%%%%%%%%%%%%%%%%%%%%%%%%%%%%%%%%%
\maketitle
\begin{abstract}
Machine learning algorithms dedicated to financial time series forecasting have gained a lot of interest. But choosing between several algorithms can be challenging, as their estimation accuracy may be unstable over time. Online aggregation of experts combine the forecasts of a finite set of models in a single approach without making any assumption about the models. In this paper, a Bernstein Online Aggregation (BOA) procedure is applied to the construction of long-short strategies built from individual stock return forecasts coming from different machine learning models. The online mixture of experts leads to attractive portfolio performances even in environments characterised by non-stationarity. The aggregation outperforms individual algorithms, offering a higher portfolio Sharpe Ratio, lower shortfall, with a similar turnover. Extensions to expert and aggregation specialisations are also proposed to improve the overall mixture on a family of portfolio evaluation metrics.
\end{abstract}

\section{Introduction}
% In their celebrated paper \cite{FamaFrench1996} propose to expand the CAPM with two fundamental factors, size and value, to explain equity returns. Since then, a massive literature proposes to extend the Fama-French framework by adding new factors, based on multiple stock characteristics. In their enumeration, \cite{GreenHandZhang2013} denote up to 330 factors proposed by the literature. From a given set of fundamental factors, a linear regression is performed to explain assets' returns. The use of too many factors could generate numerical instabilities %(QUOTE REF??)
% while the question of which factors provide independent information about the cross-section of future returns remains. \\ 
% GU: Green et al. (2013) count 330 stock-level predictive signals in published or circulated drafts. Harvey et al. (2016) study 316 “factors,” which include firm characteristics and common factors, for describing stock return behavior. They note that this is only a subset of those studied in the literature. Welch and Goyal (2008) analyze nearly 20 predictors for the aggregate market return. In both stock and aggregate return predictions, there presumably exists a much larger set of predictors that were tested but failed to predict returns and were thus never reported.

Over the last decade, data science techniques have been regularly tested in finance to improve traditional forecasting techniques. Machine learning algorithms promise, among other things, to address the challenges of high dimensional data and to consider a broader class of functions, exploiting non-linearities or interactions in the data to improve prediction. 
% These algorithms have been successfully applied for credit risk and portfolio construction. 
% For example, \cite{khandani2010consumer,butaru2016risk} use regression trees to forecast consumer delinquencies or defaults and \cite{sadhwani2021deep} consider deep neural networks for mortgage risk. 
These algorithms have been successfully applied for credit risk \citep{khandani2010consumer,butaru2016risk} and mortgage risk \citep{sadhwani2021deep}. On the portfolio construction side, \cite{MoritzZimmermann2016} use tree-based techniques to classify stock returns and build portfolios accordingly, while \cite{heaton2017deep} use deep learning hierarchical models for financial prediction and classification. Additional work focuses on time series forecasting. 
For instance, \cite{rapach2013international} explore lead-lag relationships among country stock returns and take advantage of LASSO models to forecast stock returns in the US. \cite{freyberger2020dissecting} use adaptive group LASSO to determine which firm characteristics provide
incremental information for the cross section of expected stock returns.
\cite{hutchinson1994nonparametric,yao2000option} consider a non-parametric approach with neural networks to forecast derivatives prices.
% \cite{kozak2020shrinking} use shrinkage techniques to construct a robust stochastic discount factor summarizing the joint explanatory power of a large number of cross-sectional stock return predictors.
\cite{rasekhschaffe2019machine,kozak2020shrinking} explore how machine learning models can improve stock return forecasts while avoiding over-fitting. Finally, \cite{GuKellyXiu2018} compare the performance of thirteen machine learning techniques including neural networks, random forests and linear models to forecast stock returns and build portfolios from the predictions.

In practice however, choosing a model and its hyper-parameters is not straightforward.
Initiated by \cite{bates1969combination} and based on game theory concepts (\cite{blackwell1956analog} and \cite{hannan1957approximation}),
the idea of combining predictions can be very effective for predictive learning tasks.
Averaging models may lead to a reduction in variance and induces smaller generalisation errors \citep{breiman2001random}. A key point is the diversity of the models considered in the ensemble \citep{brown2005diversity,brown2005managing}. Bagging \citep{breiman1996bagging} and Boosting \citep{freund1996experiments,schapire1990strength} are for example two popular methods for generating ensembles. 
Combining different models has also been used to improve time series forecasts, such as exponential smoothing with ARIMA \citep{bai2010forecasting}, AdaBoost with recurrent neural networks \citep{sun2018adaboost} or to forecast stock market trading patterns \citep{lin2021learning}.
\cite{weng2018predicting} predict stock prices from a neural network ensemble, a support vector ensemble, a boosted tree and a random forest.
\cite{yang2020deep}  integrate different reinforcement learning algorithms to learn a stock trading strategy. \cite{nti2020comprehensive,albuquer2022quemaking} provide a comprehensive review of ensemble techniques used in finance. Nevertheless, an algorithm may outperform others during specific time periods only and such ensemble methods are not robust to data distribution changes.

To tackle unstable accuracy over time, \cite{littlestone1994weighted} and \cite{vovk1990aggregating} independently introduced one successful approach for time series forecasting: the online aggregation of experts. 
This method allows to combine in a single approach the forecasts of a set of models, called experts \citep{cesa2006prediction}.
A new forecast is obtained with the help of sequential decision techniques and is guaranteed by the theory to be on average almost as accurate as the forecast of the best expert \citep{freund1997using,vovk1997competitive,vovk1998game}. The resulting mixture is continuously updated as soon as the expert forecasts become available. This is a desirable feature in non-stationary environments as it allows to reconsider regularly the best models. This approach is all the more appealing that it makes no assumption about the data generation process. 
The framework is also a way to meet the challenge of tuning hyper-parameters, by considering every possible parameter combinations with the same algorithm. 
In addition, aggregation with expert advice reduces the average excess risk of the estimator while benefiting from theoretically sound results on the optimal regret bound, i.e. aggregation guarantees to recover online the best possible combination of experts. 

These attractive properties partly explain why sequential aggregation procedures have been intensively studied in recent years \citep{azoury2001relative,vovk2006line,atiya2020does,petropoulos2020forecasting}. The book \cite{cesa2006prediction} provides an in-depth introduction to this approach.
Aggregation methods have been used successfully for time series forecasting applications, such as energy consumption or electricity prices \citep{GaillardGoude2014, nowotarski2018recent}, weather \citep{taillardat2016calibrated,thorey2017online}, pollution \citep{debry2014ensemble,auder2016sequential} or exchange rates \cite{amat2018fundamentals}.

In this paper, online expert aggregation is used to address the difficulty of having to choose between several investment strategies, and to ensure robustness to changing market conditions (i.e. guarantee satisfactory performance over time).
Thirteen different portfolios are constructed based on various machine learning algorithms (linear, tree-based, neural networks) forecasting one-month-ahead stock returns from firms' financial characteristics.
The dataset includes 94 characteristics (size, momentum, etc.) for a large collection of 30,000 stocks over the 1957-2016 period. Zero-net-investment portfolios are constituted based on model's forecasts, buying stocks in the the highest expected return decile and selling stocks in the lowest. Eventually, the state-of-the-art Bernstein Online Aggregation from \cite{wintenberger2017optimal} provides a convex combination of the long and short strategies based on individual experts' forecasts to build a robust portfolio. The aggregation assigns every month a weight to each expert according to its current performance. 

Results of this paper show that robust online aggregation leads to attractive portfolio performances even in adversarial environments characterised by strong non-stationarity of the data distribution. The aggregated portfolio not only outperforms the experts, but also makes the approach more robust by dynamically adapting to market changes online, which greatly reduces the shortfall risk. The aggregation allows to build an investment strategy with an annual Sharpe Ratio of 2.82, slightly higher than the best expert (a neural network) with 2.74, while having a maximum monthly loss of 7\%, more than twice as low as that same best expert (16\%). Eventually, the aggregated portfolio turnover stays close to the one of each individual strategy around 120\%.

To our knowledge, this paper provides the first application of online expert aggregation for financial strategies. This work adds on the growing literature testing machine learning techniques for portfolio management, using an adaptive mixture of long-short portfolios based on data-driven individual stock price predictions. The contribution is plural. First, state-of-the-art Bernstein Online Aggregation (BOA) is applied on portfolio construction. 
The aggregation ponders directly the stock weights of each individual expert portfolio, allowing to consider any algorithm, even black-box models. Second, the tests provide a comparison of the performance of the aggregated strategy with thirteen machine learning experts, studied by \cite{GuKellyXiu2018}, on a large dataset (more than 30,000 US stocks). Eleven of the thirteen experts used in the paper are ensemble-based by construction for stability purposes, but also to compare the performance of online aggregation with static ensemble methods. 
Finally, expert and aggregation specialisations are proposed to improve the global mixture. An expert outperforming the aggregation gives the opportunity to increase the initial set of experts with additional models derived from this best expert. In the same spirit, aggregation specialisation is introduced and explores the possibility to refine the aggregation depending on the context.

%%%%%%%%%%%%%%%%%%%%%%%%%%%%%%%%%%%%%%%%%%%%%%%%%%%%
\section{Data and Methodology}\label{sec:methodology}

Instead of relying only on one model forecast, a more robust approach considers ensemble forecasts. The aggregation framework considered in this paper tackles unstable accuracy of forecasting models in non-stationary environments in an online manner, without hypothesis on models and the data distribution.

\subsection{Expert Aggregation}\label{sec:aggregation}

A set of forecasting algorithms, called \textit{experts}, estimate independently the next value of a given sequence. 
A set of observations $\mathcal{D}_t=\{(x_1, y_1), . . . ,(x_t, y_t)\}$ is given at each time $t>0$ where the \textit{target} $y_t$ is a bounded value on $\mathbb{R}$ and $x_t\in\mathbb{R}^d$ is a feature vector
Each forecasting algorithm $k$ at $t$ is a function $f^k_t: \mathbb{R}^d \mapsto \mathbb{R}$ 
providing a forecast $f^k_t(x_{t+1})$ that has to be as close as possible to $y_{t+1}$. 
The forecasts are obtained element by element by learning the (assumed) relationship between the input space $\mathbb{R}^d$ and a bounded subset of $\mathbb{R}$.
An online expert $f^k = (f^k_0, f^k_1, f^k_2, \ldots)$ is a sequential algorithm that produces at each time $t$ an expert  $f^k_t$.

The relevance of the expert's forecast is measured at each time step by a convex loss function $\ell: \mathbb{R}\times\mathbb{R} \mapsto \mathbb{R}_+$. 
In an online setting, the goal of the experts is to minimise their cumulative empirical error 
$\sum_{t\geq0} \ell(y_{t+1}, f^k_t(x_{t+1}))$ between the true value $y_{t+1}$ and the expert's forecast $f^k_t(x_{t+1})$.

Experts aggregation is a sequential forecasting framework allowing to mix several forecasting models in a robust approach \citep{cesa2006prediction}. 
The algorithm provides as forecast a convex combination of the outcomes from a finite set of experts, where the weights are computed according to a chosen deterministic policy\footnote{Note that instead of considering convex combination of expert, some policies allow model selection aggregation problem (see \cite{cesa2006prediction,wintenberger2017optimal}).}.
% Aggregation allows to consider different classes of forecasting models in a robust and deterministic approach. 
% Aggregation with expert advice is a specific case of online learning where there are not one, but  several learners.
Let $f^1,...,f^K$ be $K$ online experts providing bounded estimations (so that losses are bounded as well).
Aggregation aims at finding the optimal online convex combination $$f_w=\sum_{k=1}^Kw_k f^k=\left(\sum_{k=1}^Kw_{k,t}f^k_t\right)_{t\geq0}$$ with weights $w_{k,t}\in S$ where $S$ is a closed and bounded subset of $\mathbb{R}^K$. In the following $S=\{w_k\in\mathbb{R}^K_+, \sum_{k=1}^Kw_k=1\}$.
%, but other closed and bounded sets could be used according to different aggregation rules, for example a ball of radius $r$ in $\mathbb{R}^K$ $S=\mathbb{B}_K(r)$.
% \geq 0$ and $\sum_{k=1}^Kw_{k,t}=1$.
% $S=\{w\in \mathbb{R}^K_+, \sum_{i=1}^K w_i=1\}$
The performance of the online aggregation procedure is measured by the cumulative error 
$\sum_{t\geq0}\ell(y_{t+1}, f_{w,t}(x_{t+1}))$  between the target $y_{t+1}$ to be predicted and the mixture's forecast $f_{w,t}(x_{t+1})$. 

However, if the experts' accuracy is low, so will be the accuracy of the mix, it is thus impossible to ensure a low cumulative loss for the aggregation in absolute terms. Therefore, the aggregation framework seeks to ensure a low cumulative error compared to the cumulative errors of the experts. To do so, the mixture is compared to the best possible fixed combination of experts, called \textit{oracle}\footnote{Specific aggregations or settings include non-stationary oracles, see for instance \cite{herbster1998tracking}.}. The goal of the aggregation is thus to retrieve online the oracle.

The \textit{regret} \citep{freund1997using} compares a given online aggregation procedure with the oracle in terms of cumulative errors. 
The regret is defined at time $T$ by
\begin{eqnarray}
R_T = \sum_{t=0}^{T} \ell(y_{t+1}, f_{w,t}(x_{t+1})) - \inf_{u \in S}\left\{\sum_{t=0}^{T}\ell(y_{t+1},f_{u,t}(x_{t+1}))\right\}, \label{eq:regret}\nonumber
\end{eqnarray}
where the first term is the cumulative error of the mixture and the second term is the approximation error, i.e. the cumulative error of the oracle compared to the target $y_{t+1}$ . %where the first term is the average error of the mixture and the second term is the approximation error.
By minimising regret, one seeks to avoid sub-optimal mixtures and thus reduce the number of actions taken where, in hindsight, a better choice would have been possible. 
% The policy that directs these choices to minimise regret focuses solely on how the experts' weights are allocated.
These choices are directed by a given policy, called \textit{rule}.

The rule of an aggregation determines how the weights are assigned to each expert.
The rules are deterministic, need all expert forecasts at each time\footnote{\cite{devaine2013forecasting, gaillard2014second} propose theoretical guarantees about regret convergences when some expert predictions are missing. Called sleeping experts, the missing estimates can be replaced by those from the aggregation. \cite{mourtada2017efficient} explore aggregations of a set of experts that is no longer fixed but can increase over time, which is particularly useful for dealing with non-stationarity.} and depend on a learning rate parameter $\eta>0$.
The learning rate, which is preferably tuned online, guides the aggregation rule adaptability to the environment. % \cite{devaine2013forecasting}
Having a high $\eta$ leads to follow the best expert, while a lower rate leads the mixture to a more uniform and conservative distribution.
Many rules exist in the literature and differ according to the application.
\cite{littlestone1994weighted} and \cite{vovk1990aggregating} propose to use an online convex aggregation rule called Exponentially Weighted Average (EWA) allowing rough changes in the weights allocation. 
Multiple Learning rate (ML Poly, \cite{cesa2003potential,gaillard2014second}) has its own learning parameter calibration rule which is faster than the empirical tuning described by \cite{devaine2013forecasting}. 
Fixed Share forecaster (FS, \cite{herbster1998tracking}) competes not only with the best fixed expert but also with the best sequence of experts and Ridge allows non-positive weights and non-convex combinations \citep{azoury2001relative,vovk2006line}.
Most of the aggregation rules, in particular the one used in this paper, ensure that the regret converges to zero when $T$ goes to infinity.
So, this study focuses on reducing the approximation error by increasing the heterogeneity of the expert set (see \cite{GaillardGoude2014} and \cite{Stoltz2005} for further details).

This paper considers the Bernstein Online Aggregation (BOA, \cite{wintenberger2017optimal}).
At time $t$, BOA assigns a new weight to each expert according to its accuracy compared to the other experts, by minimising the loss 
$\ell_{k,t+1} = \ell(y_{t+1} , f_{k,t}(x_{t+1})) - \ell(y_{t+1} , f_{w,t}(x_{t+1}))$.
Given the losses $\ell_t=(\ell_{1,t},\dots,\ell_{K,t})$ suffered by the experts at each instance $t$, BOA assigns to expert $k$ the weight

$$
w_{k,t} = \frac{\exp(-\eta \ell_{k,t}(1+\eta \ell_{k,t}))}{\exp(-\eta \ell_{w,t}(1+\eta \ell_{w,t}))}w_{k,t-1}, 
$$ %\label{eq:weight}

where $\ell_{w,t}=\sum_{k=1}^K w_{k,t-1} \ell_{k,t}$ is the loss suffered at time $t$ by the aggregation at time $t-1$ (with weights $w_{t-1}$).
The second order refinement of the loss is designed to penalise large errors and stabilises the weight allocation.
BOA benefits from a faster rate of convergence than other rules. 
The learning rate $\eta$ is optimally tuned in the BOA process and ensures minimising regret with the fast rate of convergence $\log(K)/T$ in deviation. 
BOA procedure is reported in Algorithm \ref{alg:boarule}.
For further details, see \cite{wintenberger2017optimal}. 

\begin{center}
\begin{minipage}{.55\linewidth}
\begin{algorithm}[H]
\caption{Bernstein Online Aggregation (BOA)}
\begin{algorithmic}
\REQUIRE weights  $w_{k,0}>0$ s.t. $\sum^K_{k=1} w_{k,0} =1 $, learning rate $\eta>0$
\FOR{each $t=1,...,T$}
\FOR{each $k=1,...,K$}
\STATE $$w_{k,t} = \frac{\exp(-\eta \ell_{k,t}(1+\eta \ell_{k,t}))}{\exp(-\eta \ell_{w,t}(1+\eta \ell_{w,t}))}w_{k,t-1}$$
\ENDFOR
\ENDFOR
\end{algorithmic}
\label{alg:boarule}
\end{algorithm}
\end{minipage}
\end{center}

\FloatBarrier
%%%%%%%%%%%%%%%%
\subsection{Application to Financial Portfolio}

The online aggregation procedure is applied to portfolio returns, by minimising the cumulative loss between each expert’s portfolio and the best possible portfolio, the target.\footnote{Note that an alternative approach would have been to aggregate stock return forecasts rather than portfolios' weights. But, because the final goal is to improve directly portfolio performance from any (potentially black-box) strategy and not return forecast accuracy, the online mixture is applied on portfolio weights.} The target for the long (resp. short) portfolio is obtained by buying at each rebalancing date the 10\% best (resp. worst) performing stocks.
In practice, the aggregation assigns at the end of each month a vector of weights to the expert portfolios to minimise the difference between the mixture returns and this optimal portfolio returns. 
Note that the target should not be confused with the oracle discussed in Section \ref{sec:aggregation}, which corresponds to the best possible mixture of experts (and could be very different from the target depending on the quality of the experts). 
Two aggregations are applied, one for the long strategies and another one for the short strategies. The best long experts are not necessarily the same as the short ones at each instant. Using two distinct aggregations allows to take advantage of different experts at different times.
The returns of the long-short aggregation are the difference between the long aggregation portfolio and the returns of the short one.

\subsection{Data}\label{sec:data}

Data comes from Wharton Research Data Services (\cite{WRDSdatabase}), including CRSP and Compustat database, and covers more than 30,000 US stocks over 1957-2017 period. 
The 94 standard firm characteristics used by \cite{GuKellyXiu2018} are considered as features to feed the stock return forecasting algorithms.\footnote{In the original paper, 920 stock characteristics are used. For simplicity, the eight macroeconomic predictors as well as the interactions between firm-level characteristics and macroeconomic state variables are omitted. However, the benchmark results obtained in this paper are essentially the same and detailed in Table \ref{tab:benchmark} in Appendix.}
Twenty among these features are updated monthly, thirteen updated quarterly and sixty-one updated annually. 
Following \cite{GuKellyXiu2018} and \cite{freyberger2020dissecting}, a cross-section rank transformation is performed each month on all firm characteristics that maps these ranks into the $[-1,1]$ interval.
Missing data are replaced by their cross-sectional median.
In order to avoid forward-looking bias (information at month $t$ is only known at month $t+1$ for monthly characteristics, $t+4$ for quarterly and $t+6$ for annual ones), monthly variables are delayed by one month, quarterly data by four months, and annual data by six months.

\FloatBarrier
%%%%%%%%%%%%%%%%%%%%%%%%%%%%%%%%%%%%%%%%%%%%%%%%%%%
\section{Expert Portfolio}\label{sec:empiricalres}

The online aggregation is compared to the strategies based on single forecasting algorithms and follow the same methodology as \cite{GuKellyXiu2018} regarding data construction, forecasting windows and models construction. This section describes the datasets and the thirteen forecasting models. Then, the performance of long-short strategies based on individual experts is presented.

\subsection{Forecasting Models}\label{sec:empirical_gu_descrip}

Stocks are characterised by a set of features, such as firm size or stock return momentum. For each month and each stock, a forecasting model is fed with these features to predict next month return. 
Due to computationally intensive forecasting procedures, the models are re-calibrated only each year. The training set size starts in 1957 with 18 years and increases with time. Models are re-fitted by increasing the training sample by one year. The validation set size of 12 years is maintained constant by rolling it forward to include the most recent year. The unobserved one-year testing set is picked within a 30-year period from 1987 to 2016. So, the first training (over 30 ones) is done on the 1957-1974 period, the validation on 1975-1986 and the out-of-sample test on year 1987. The second training is then done on 1957-1975, validation on 1976-1987, test on 1988, and so on until the testing year reaches 2016.

A unique model per method is trained for all stocks, and the model stays the same over the training period as done in \cite{GuKellyXiu2018}.
This avoids intensive computational costs and tends to stabilise return estimates of individual stocks. Note that the forecasting models are not strictly online, but their associated portfolios are, consistent with the online aggregation framework. The thirteen models are reported in Table \ref{tab:expertlist} and their hyper-parameters in Table \ref{tab:hyperparameters} in Appendix.

\begin{table}[H]
\centering
\begin{tabular}{ll}
\toprule
\textbf{Family} &\textbf{Model} \\ 
\midrule
\textbf{Linear }& Ordinary Least Square (OLS+H)  \\ 
& Ordinary Least Square 3 factors (OLS3+H) \\
& Generalised Linear Model with group Lasso (GLM+H) \\
& Elastic Net (ENet+H) \\ 
\hline
\textbf{Linear with}  & Partial Least Square (PLS) \\ 
\textbf{dimension reduction }& Principal Component Regressor (PCR)\\ 
\hline
\textbf{Tree-based} & Random Forest (RF)\\
& Gradient Boosting Regressor Tree (GBRT+H) \\
\hline
\textbf{Neural network} & Neural Networks (NN1-NN5) \\
\bottomrule
\end{tabular}
\caption{Forecasting Model Description.}
\label{tab:expertlist}
\end{table}
Models are trained by minimising the squared error between the observed and the estimated stock returns,
except for models followed by ``+H'' indicating the use of Huber loss defined in Appendix \ref{sec:metrics}. Huber loss minimises the squared loss when residuals are below an (optimised) threshold and the absolute loss above. Huber is thus robust to outliers while not ignoring their effects.

Except for Partial Least Square (PLS) and Principal Component Regressor (PCR), all the models are ensemble-based by construction. 
A given model is trained several times on the dataset, then averaged to produce the forecast.
This allows to stabilise the estimators, reducing over-fitting for the linear models with Huber loss and the neural networks trained using gradient descent, but also to compare aggregation with static ensemble methods.

The performances of the forecasting models are reported in Table \ref{tab:r2scores} in Appendix. 
Following \cite{lewellen2014cross}, the accuracy of three basic forecasting benchmarks is shown in Table \ref{tab:benchmark} in Appendix, based on three ordinary least squared regressions with respectively 3, 7 or 15 variables.

% \FloatBarrier
%%%%%%%%%%%%%%%%%%%%%%%%%%%%
\subsection{Performance of Experts' Portfolios}\label{sec:ptfstrat}

Stocks are sorted according to each model's predictions. A long (resp. short) portfolio is built by buying 10\% of the stocks having the highest (resp. lowest) estimated returns.
Both equally and value weighted portfolios are considered.
Value weighted results are presented in the appendix.
Value weighted portfolios are less sensitive to illiquidity of small cap stocks, but as the objective functions of both forecasting models and aggregation minimize equally weighted forecast errors, the main paper focuses on equally weighted portfolios.
When comparing portfolio performance with the estimation accuracy in Table \ref{tab:r2scores}, improved forecasts do not necessarily lead to better portfolios. For instance, the portfolio OLS+H outperforms the experts based on linear and tree algorithms, while having one of the lowest \%R2 among the forecasting models. 
Average monthly returns generally increase monotonically decile by decile for each algorithm, as reported in Appendix Table \ref{tab:decile_equally} for equally weighted portfolios and Table \ref{tab:decile_value} for value weighted. 

Table \ref{tab:ptfperf} reports the performances of each expert portfolio on the out-of-sample test period. The performance metrics definitions are precised in Appendix \ref{sec:metrics}. 
In general, expert performance improves with the model complexity. An exception occurs with the linear model OLS+H which provides similar performance to the neural networks\footnote{This is a notable distinction from the results of \cite{GuKellyXiu2018} where tree-based models and neural networks significantly outperformed OLS+H.}.
Portfolio performances are in line with forecast accuracy results. Neural network offer the best portfolio performance, outperforming the other strategies with an annual Sharpe Ratio (SR) always greater than 2.2. 
The expert NN2 dominates the other algorithms with an annual return of 0.5 for the equally weighted portfolio and 0.37 for value weighted, leading to a SR of 2.74 and 2.67 respectively. Linear models ENet+H and GLM+H are also valuable strategies, with a SR of 1.77 and 1.81 respectively, very close to experts based on dimension reduction (PLS 1.85 and PCR 1.78). 
Tree-based models GBRT+H and RF have a SR of 1.71 and 1.96 respectively, thanks to their low volatility (0.15 and 0.14 respectively). Surprisingly, OLS+H proposes a comparable SR (2.28) as neural networks. While the model is restricted to linear functions and not particularly well fitted for high-dimensional data, when considering long-short stock strategies, OLS+H competes with more sophisticated models able to deal with non-linear relationships between variables. OLS3+H (which has a limited number of stock characteristics) has the lowest SR at 1.1.

\begin{table}[H]
    \begin{center}
    \resizebox{\textwidth}{!}{
    \begin{tabular}{l cccc cccc cccc c}
    \toprule
    & OLS&OLS3&PLS&PCR&ENet&GLM&RF&GBRT&NN1&NN2&NN3&NN4&NN5\\
    &+H&+H&&&+H&+H&&+H&&&&&\\
    \midrule
    Ann Ret&0.36&0.21&0.31&0.30&0.30&0.32&0.28&0.25&0.48&\textbf{0.50}&0.48&0.48&0.43\\
    % Cum. Ret.&10.76&6.32&9.24&9.01&9.09&9.57&8.31&7.60&14.54&\textbf{15.11}&14.43&14.51&13.02\\
    Vol&0.16&0.19&0.17&0.17&0.17&0.18&\textbf{0.14}&0.15&0.20&0.18&0.20&0.19&0.20\\
    SR&2.28&1.11&1.85&1.78&1.77&1.81&1.96&1.71&2.39&\textbf{2.74}&2.42&2.56&2.21\\
    Skew&0.52&0.77&0.14&0.46&-0.05&-0.26&0.97&1.72&2.18&2.27&1.69&1.75&\textbf{2.40}\\
    Kurt&\textbf{4.36}&17.89&7.96&7.50&7.42&8.63&7.03&12.63&19.44&13.94&11.12&11.47&19.56\\
    Max DD&0.25&0.63&0.44&0.31&0.40&0.40&0.23&0.24&0.28&\textbf{0.17}&0.29&0.23&0.47\\
    Max Loss&\textbf{0.13}&0.36&0.23&0.20&0.22&0.27&0.17&0.16&0.28&0.16&0.23&0.22&0.23\\
    Turnover&1.26&1.50&1.15&1.27&1.28&1.36&\textbf{0.92}&1.25&1.24&1.23&1.20&1.20&1.15\\
    \bottomrule
    \end{tabular}
    }
    \caption{Performance of Experts' Portfolios}
    \label{tab:ptfperf}
    \end{center}
    Note: \textit{This table presents the performance of long-short strategies based on individual forecasting models. Columns Ann Ret, Vol, Skew, Kurt, SR, Max DD, Max Loss and Turnover stand for annualised average return, volatility, skewness, kurtosis, annual Sharpe Ratio, maximum drawdown,  1-month maximum loss and portfolio turnover. The metrics are computed on the test period 1987-2016. Portfolios are equally weighted.}
\end{table}

At first sight, selecting a best strategy is not obvious. NN2 offers better returns and SR, but the low turnover of RF (92\% against an average of 120\%) or the low maximum monthly loss of OLS+H (13\%) makes these strategies appealing in practice.
NN5, although appealing in terms of SR, is characterised by higher extreme risks (highest kurtosis among all experts and a maximum drawdown substantially greater than that of NN2 (47\% against 17\%).
The qualitative conclusions remain when considering value weighted portfolios (reported in Table \ref{tab:ptfperf_value} in Appendix).

% \FloatBarrier
%%%%%%%%%%%%%%%%%%%%%%%%%%%%
\section{Aggregation of Portfolios}\label{sec:stratagg}

This section shows how the aggregation of individual strategies based on machine learning models can enhance portfolio performance and adapt to changing market conditions. Variant aggregations that improve the mixtures in specific contexts are presented. Finally, the importance of each expert in the aggregation is examined and expert specialisation is discussed.
% \\
% \red{Let's point out that we aggregate the strategies and not the return forecasts as the objective is to maximize portfolio strategy's performances. As provided in Appendix C, forecast improvement does not necessarily leads to better portfolios:  for instance, OLS+H leads to better portfolio's performances than other linear and tree-based models while having a lower \%R2.}

\subsection{Aggregated Portfolio Performance}\label{sec:longshortagg}

\begin{table}[H]
    \begin{center}
    \resizebox{\textwidth}{!}{
    \begin{tabular}{l ccc ccc}
    \toprule
    &Best Expert&\multicolumn{2}{c}{Fixed Combination}&\multicolumn{2}{c}{Adaptative Mixture}& Oracle\\
    \cmidrule(lr){3-4} \cmidrule(lr){5-6}
    &NN2&PtfUNI&Best Convex&Best Convex&PtfBOA&Best Convex\\
    &&&on Valid. Set&One-year Rolling&&\\
    \midrule
    Ann Ret&\textbf{0.50}&0.36&0.36&0.43&0.49&0.50\\
    % Cum. Ret.&\textbf{14.36}&10.45&10.25&12.24&14.03&14.38\\
    Vol&0.18&\textbf{0.14}&0.16&0.16&0.18&0.17\\
    SR&2.74&2.56&2.28&2.60&\textbf{2.77}&2.92\\
    Skew&2.27&1.19&0.52&1.65&\textbf{3.11}&2.90\\
    Kurt&13.94&10.17&\textbf{4.36}&10.58&19.63&17.93\\
    Max DD&0.17&0.24&0.25&0.17&\textbf{0.08}&0.07\\
    Max Loss&0.16&0.18&0.13&0.16&\textbf{0.08}&0.07\\
    Turnover&1.23&\textbf{1.22}&1.26&\textbf{1.22}&1.23&1.24\\
    \bottomrule
    \end{tabular}
    }
    \end{center}
    \caption{Performance of Aggregated Portfolios.}
Note: \textit{Columns Ann Ret, Vol, Skew, Kurt, SR, Max DD, Max Loss and Turnover stand for annualised average return, volatility, skewness, kurtosis, annual Sharpe Ratio, maximum drawdown,  1-month maximum loss and portfolio turnover. The metrics are computed on the test period 1987-2016. 
Expect for NN2, all the portfolios are convex combinations of the thirteen experts. 
The best convex combination on the validation set is a fixed combination calibrated on 1986. The best convex one-year rolling mixture assigns the best convex combination of experts estimated the previous year to the next year. The oracle is the best possible convex mixture on the test period, unachievable in practice. Portfolios are equally weighted.  }
\label{tab:stratstatdesc}
\end{table}

BOA rule is applied on the expert portfolios, minimising the square loss between the best possible portfolio returns and the returns of the $K=13$ expert strategies listed in Table \ref{tab:expertlist}.
The resulting portfolio is called PtfBOA. By dynamically weighting strategies, one can expect to retrieve (at least) the best expert's portfolio returns and reduce the risk of betting on only one expert. A uniform mixture of the $K$ portfolios, called PtfUNI, is used as a benchmark and assigns a constant weight of $1/K$ to each expert throughout the test period.

Table \ref{tab:stratstatdesc} shows the best expert in terms of annual Sharpe Ratio (NN2), PtfUNI, PtfBOA and the oracle, i.e. the best possible convex combination over the test period, unachievable in practice. 
Two additional mixtures are presented to compare BOA with simple ensemble approaches. 
The best convex combination calibrated over the last year of the validation set provides a fixed weighting that is less naive than the uniform mixture.
To adapt to changing market conditions, the best one-year rolling convex mixture assigns the best fixed convex combination estimated the previous year to the next year. The latter does not benefit from the same theoretical guarantees than BOA and can induce rough variations in portfolio weights through time.

The aggregation PtfBOA brings a significant improvement to classical machine learning techniques and the different mixtures.
BOA portfolio provides the highest SR at 2.77, followed by NN2 (2.74), while decreasing substantially the portfolio maximum monthly loss (8\%) compared to NN2 and all portfolios. 
PtfBOA has the highest skewness (3.11) and the lowest max DD (8\%), with a similar turnover (123\%) as NN2 and the mixtures.
These results are all the more encouraging that the expert aggregation does not directly optimise these indicators, but only considers the error between the monthly returns of the expert portfolios and the target. 
The uniform aggregation PtfUNI offers the third best SR at 2.56, ex-aequo with NN4, despite lower annalised returns (0.36) compared to NN2 and PtfBOA. 
Value weighted strategies give similar results (see Table \ref{tab:stratstatdesc_value} in Appendix), PtfBOA being able to outperform the experts and mixtures in multiple metrics. 
As expected, the turnover of all value weighted portfolios is lower than in the equally weighted case. 

The oracle's SR (2.92) indicates that there is only marginal room for improvement, by designing better objective functions or better rules for the online aggregation.
The two simple ensemble-based portfolios (the fixed and one-year rolling best convex combination), with a SR of 2.28 and 2.60 respectively, do not manage to beat the best expert. Both mixtures also underperform PtfBOA in terms of returns, SR, max DD and max Loss.
To put emphasis on the relationship between experts and the online aggregation, the following analysis focuses on PtfBOA and PtfUNI.

% \FloatBarrier
\subsection{Mixture Analysis}

\paragraph*{Weights analysis}  

\begin{figure}[h]
  \begin{center}
  \includegraphics[width=\textwidth]{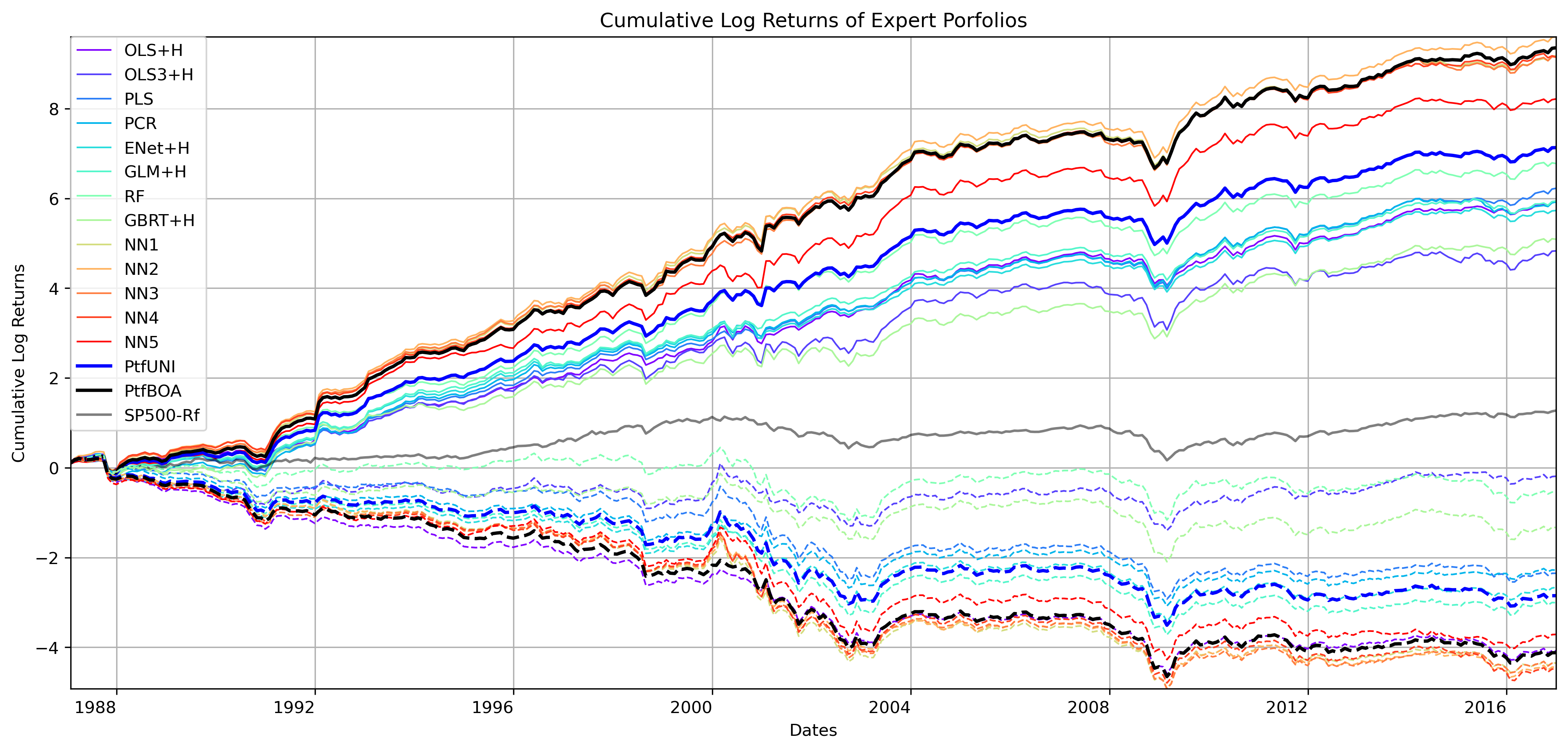}
  \includegraphics[width=\textwidth]{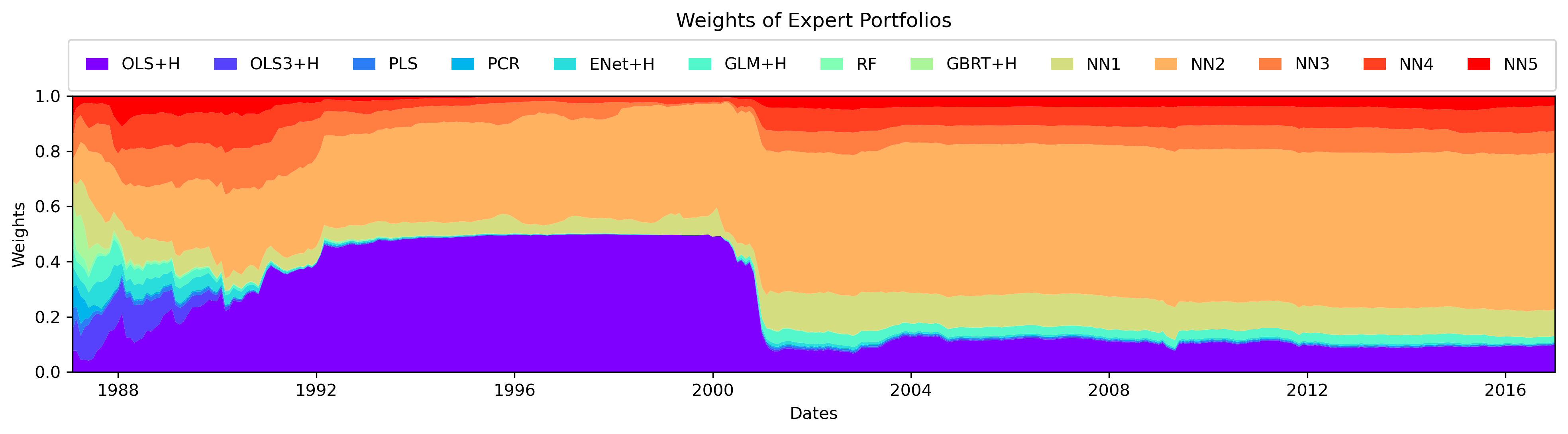}
  \end{center}
  \vspace{-0.5cm}
  \caption{Cumulative Returns of Portfolios and Experts' Weights of PftBOA.}
  Note: \textit{The first graph presents cumulative log returns of portfolios on the test period as well as S\&P500 (in gray). Full lines (resp. dash lines) indicate long positions (resp. short). Black bold lines correspond to aggregation PtfBOA and blue bold lines to the uniform mixture PtfUNI. Both aggregations include the 13 strategies OLS+H, OLS3+H, PLS, PCR, ENet+H, GLM+H, RF, GBRT+H, NN1, NN2, NN3, NN4, and NN5. The bottom graph plots the average expert weights of the long and short aggregations of PtfBOA over the period. Portfolios are equally weighted.}
  \label{fig:boa}
\end{figure}

Figure \ref{fig:boa} displays the dynamic weights of the thirteen experts within BOA as well as the cumulative log returns of both long and short strategies. 
The bottom graph highlights how aggregation proceeds: starting with uniform weights, a transition phase over the first few months gives way to the best experts until the mixture converges. 

The mixture favors neural networks and OLS+H. 
NN2 is the best expert all over the test period for the long strategy in terms of cumulative returns.
The linear model OLS+H is the best expert for the short strategy before 2001, and therefore share a large part of the weighting in the mixture (aroung 40\%).
In 2001, a regime break (coinciding with the dot.com bubble burst) reduces OLS+H importance in the mixture at the benefit of neural networks.
Aggregation adapts quickly its weights during the 2001 regime shifts, which partly explains the attractive performance and robustness of the strategy.
PtfBOA is the second best profitable strategy, for both the long and short portfolios.  
It is worth noting that the difference in cumulative performance between PtfBOA and PtfUNI is large, and encourages the adoption of an online mixture.

Over the entire test period, expert weights are relatively stable from 1992 to 2000 and then from 2001 until 2016. 
These two regimes of stable outperformance of the best experts encourage the mix to ``follow the leader'' instead of considering a more heterogeneous mixture.
This is an interesting result, especially in a non-stationary environment where the most profitable strategy may vary from one instant to the next. Here, aggregation is based mainly on the best portfolio and benefits little from the opinion of the other experts.

Interestingly, during the 2008 Subprime crisis, all experts and the aggregation experience a large drop in performances. However, the crisis has only a small impact on individual experts' weights in the aggregation. Neural networks manage to get higher average returns compared to the other experts on this specific period, and thus retain their dominant position in the mix (Figure \ref{fig:aretheatmap} in Appendix). 

Average weights of the experts on the test period are given in the Table \ref{tab:weights} in Appendix. In particular, NN2 and OLS+H share 67\% of the weight allocated to the thirteen strategies on average over the test period. PtfBOA has the closest weight combination to the oracle.

\begin{figure}[H]
    \begin{center}
    \includegraphics[width=.8\textwidth]{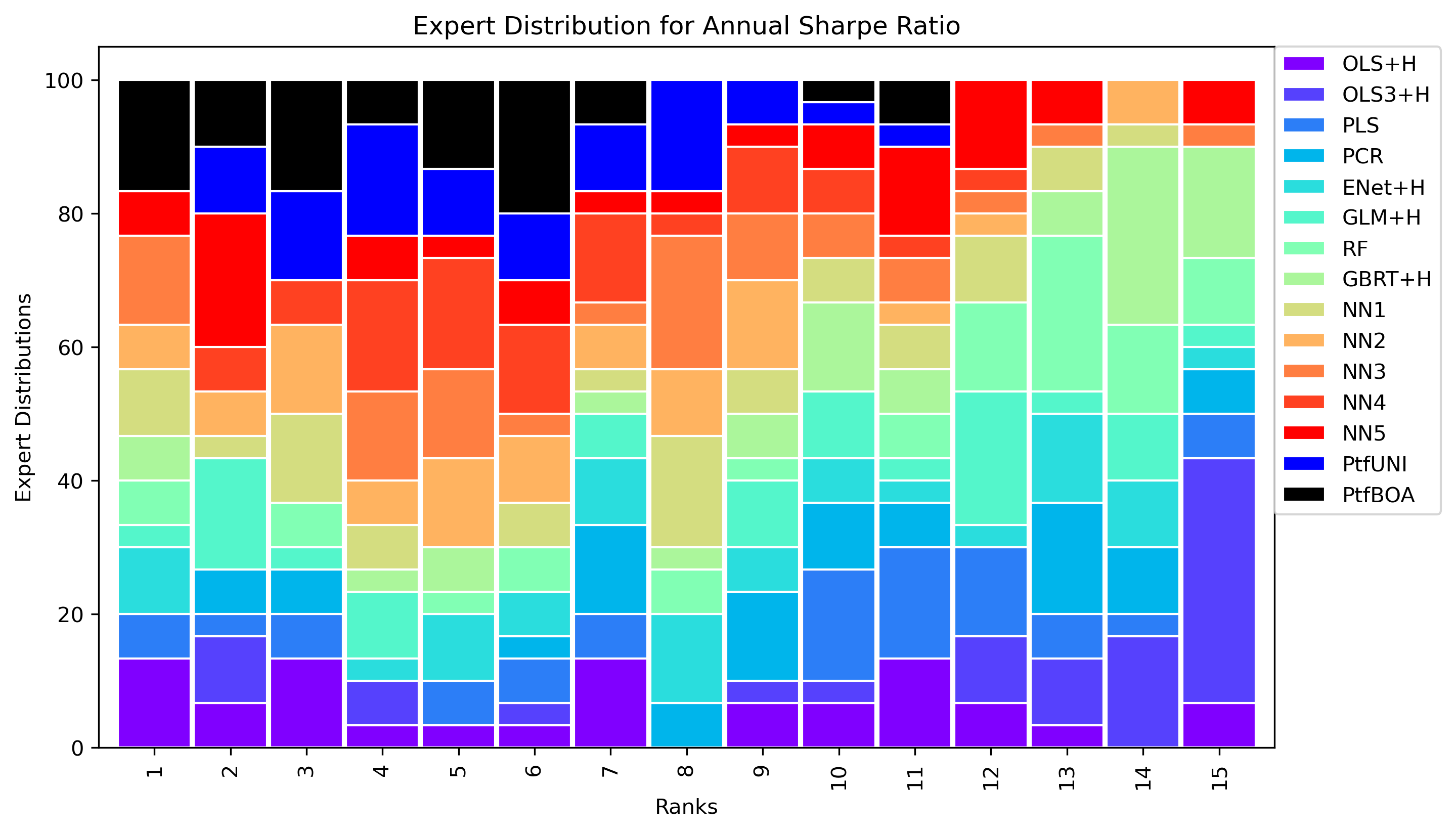}
    \end{center}
    \vspace{-0.5cm}
    \caption{Experts' Sharpe Ratio Ranks Distribution.}
    Note: \textit{This graph presents the distribution of each annual Sharpe Ratio's rank for the thirteen individual experts, the uniform mixture and the two aggregations (PtfBOA and PtfUNI). The distribution is obtained by counting the number of times an expert gets the best, second (and so on) Sharpe Ratio. Portfolios are equally weighted.}
    \label{fig:aggranks}
\end{figure}

\paragraph*{Experts ranking} Figure \ref{fig:aggranks} illustrates the distribution of each Sharpe Ratio's rank for the thirteen individual experts, the uniform mixture and the two aggregations (PtfBOA and PtfUNI). Each column indicates the number of time that the strategy has been ranked 1, 2, 3, $\dots$, 15 in terms of SR over the test period (1987-2016). 
PtfBOA reaches rank 1 close to 17\% of the time over the test period, followed by OLS+H (13\%). However, OLS+H is often one of the least profitable experts, as illustrated by its significant proportion among low ranks (from 9 to 15). 
The five neural networks (over thirteen experts) and PtfBOA represent more than half of the overall area on the top 3 ranks. 
In particular, PtfBOA appears in 14\% of the top 3 SR, when OLS+H reaches 11\% and NN2 9\%, close to the uniform mixture PtfUNI 8\%. 
Besides, the ranks of NN2 or OLS+H are more disparate compared to the ranks of the aggregation.
%Despite illustrating the distribution of experts for each rank, note that the figure does not tell how close the Sharpe Ratios are between two ranks.

\subsection{Sub-samples Analysis}
\label{sec:aggsubportfolio}

% In the previous analysis, the complete universe of US stocks is considered, including small and micro caps, that may be more difficult to rebalance frequently due to liquidity constrains.
In this section, the aggregation is performed on two sub-samples of US stocks, specifically the top and bottom 1000 stocks in terms of market capitalisation.
The aim is to test if the aggregation performance hold on the sub-sample of large stocks, because small and micro cap stocks may be more difficult to trade frequently due to liquidity issues. 
% \textcolor{red}{ajouter SMB, HML et tout le tralala}.

\begin{table}[H]
\begin{center}
\resizebox{\textwidth}{!}{
\begin{tabular}{l cccc cccc ccccc cc}
\toprule
&OLS&OLS3&PLS&PCR&ENet&GLM&RF&GBRT&NN1&NN2&NN3&NN4&NN5&PtfUNI&PtfBOA\\
&+H&+H&&&+H&+H&&+H&&&&&&&\\
\midrule
&\multicolumn{15}{c}{Top 1000 Market Caps}\\
Ann Ret&0.15&0.06&0.13&0.13&0.06&0.06&0.08&0.05&0.16&\textbf{0.17}&0.15&0.15&0.11&0.11&0.13\\
Vol&0.16&0.18&0.17&0.16&0.18&0.18&0.16&\textbf{0.14}&0.23&0.22&0.24&0.23&0.21&\textbf{0.14}&\textbf{0.14}\\
    SR&0.94&0.34&0.79&0.78&0.34&0.34&0.52&0.39&0.70&0.76&0.64&0.65&0.54&0.82&\textbf{0.95}\\
    &\multicolumn{15}{c}{Bottom 1000 Market Caps}\\
Ann Ret&0.48&0.58&0.46&0.47&0.58&0.58&0.83&0.73&0.97&\textbf{1.01}&0.97&0.96&0.87&0.73&0.97\\
Vol&\textbf{0.21}&0.26&0.23&0.23&0.26&0.26&0.34&0.32&0.38&0.40&0.39&0.38&0.42&0.24&0.37\\
    SR&2.25&2.21&1.99&1.99&2.21&2.21&2.43&2.26&2.57&2.56&2.51&2.5&2.09&\textbf{3.07}&2.59\\
\bottomrule
\end{tabular}
}
\end{center}
\caption{Statistical Performance of Sub-portfolios: Small and Large Caps.}
Note: \textit{This table presents the performance of individual experts portfolios, PtfUNI and PtfBOA on two different sub-samples: the top 1000 and the bottom 1000 stocks in terms of market capitalisation. Columns Ann Ret, Vol, SR report annualised average return, volatility and annual Sharpe Ratio. The metrics are computed on the test period 1987-2016. Portfolios are equally weighed. }
\label{tab:substats}
\end{table}

Table \ref{tab:substats} reports the Sharpe Ratios of experts and aggregation-based strategies on the two different market capitalisation universes.
It is striking to see that the best experts are very different on the two universes of stocks. OLS+H provides the highest SR (0.94) among the largest stocks (top 1000), while NN1 is the best expert on small stocks (bottom 1000) with a SR at 2.57, very close to NN2 at 2.56.
However, PtfBOA outperforms all individual experts for the universe of large stocks, with a SR of 0.95, and an attractive SR (2.59) on the smallest ones. The naive constant weighting PtfUNI provides the highest SR for small stocks (3.07). 
% \red{This could be explained by the highest volatility of little market capitalisations which can pull down some experts, where the uniform mixture muffles the portfolio return losses.}
Depending on the stock size, best experts differ and the online aggregation allows to adjust experts' weights accordingly.

% %%%%%%%%%%%%%%%%%%%%%%%%%%%%%%%%%%%%%%%%%%%%%%%%%%%%%%%%%%%%%%%%%
\subsection{Improving Aggregation}

A strong advantage of the aggregation lies also in its easy adaptation to alternative objectives.
Two improvements of the aggregation are proposed, either by adding some prior in the global mixture (Section \ref{sec:pretrained}) or by specialising the best experts (Section \ref{sec:expertspecialisation}).

\subsubsection{Pre-trained Aggregation}
\label{sec:pretrained}
At initialisation, expert's weights are uniformly set (at 1/K with $K$=13 the number of experts) and are then updated according to their losses and BOA rule. In order to speed up the convergence, one can pre-train the online mixture on the last year of the validation set, namely 1986. Thus, over the year 1987, the aggregation benefits from a prior.

\begin{table}
    \begin{center}
            \begin{tabular}{lcccc}
    \toprule
    &Best Expert (NN2)&PtfUNI&Original BOA &Pre-trained BOA\\
    \midrule
    Ann Ret&\textbf{0.50}&0.36&0.49&0.49\\
    % Cum. Ret.&\textbf{15.11}&10.64&14.72&14.79\\
    Vol&0.18&\textbf{0.14}&0.18&0.17\\
    SR &2.74&2.56&2.77&\textbf{2.78}\\
    Skew     &2.27&1.22&3.11&\textbf{3.15}\\
    Kurt     &13.94&\textbf{10.49}&19.63&19.66\\
    Max DD   &0.17&0.24&\textbf{0.08}&\textbf{0.08}\\
    Max Loss &0.16&0.18&\textbf{0.08}&\textbf{0.08}\\
    Turnover &1.23&\textbf{1.22}&1.23&1.23\\
    \bottomrule
    \end{tabular}
    \end{center}
    \caption{Pre-trained Aggregation Portfolio Performance.}
    \label{tab:validagg}
    Note: \textit{Columns Ann Ret, Vol, SR, Skew, Kurt,  Max DD, Max Loss and Turnover stand for annualised average return, annualised volatility, annual Sharpe Ratio, skewness, kurtosis,  maximum drawdown,  1-month maximum loss and portfolio turnover. Metrics are computed on the test period. Portfolios are equally weighted. }
\end{table}
\begin{figure}
    \begin{center}
    \includegraphics[width=\textwidth]{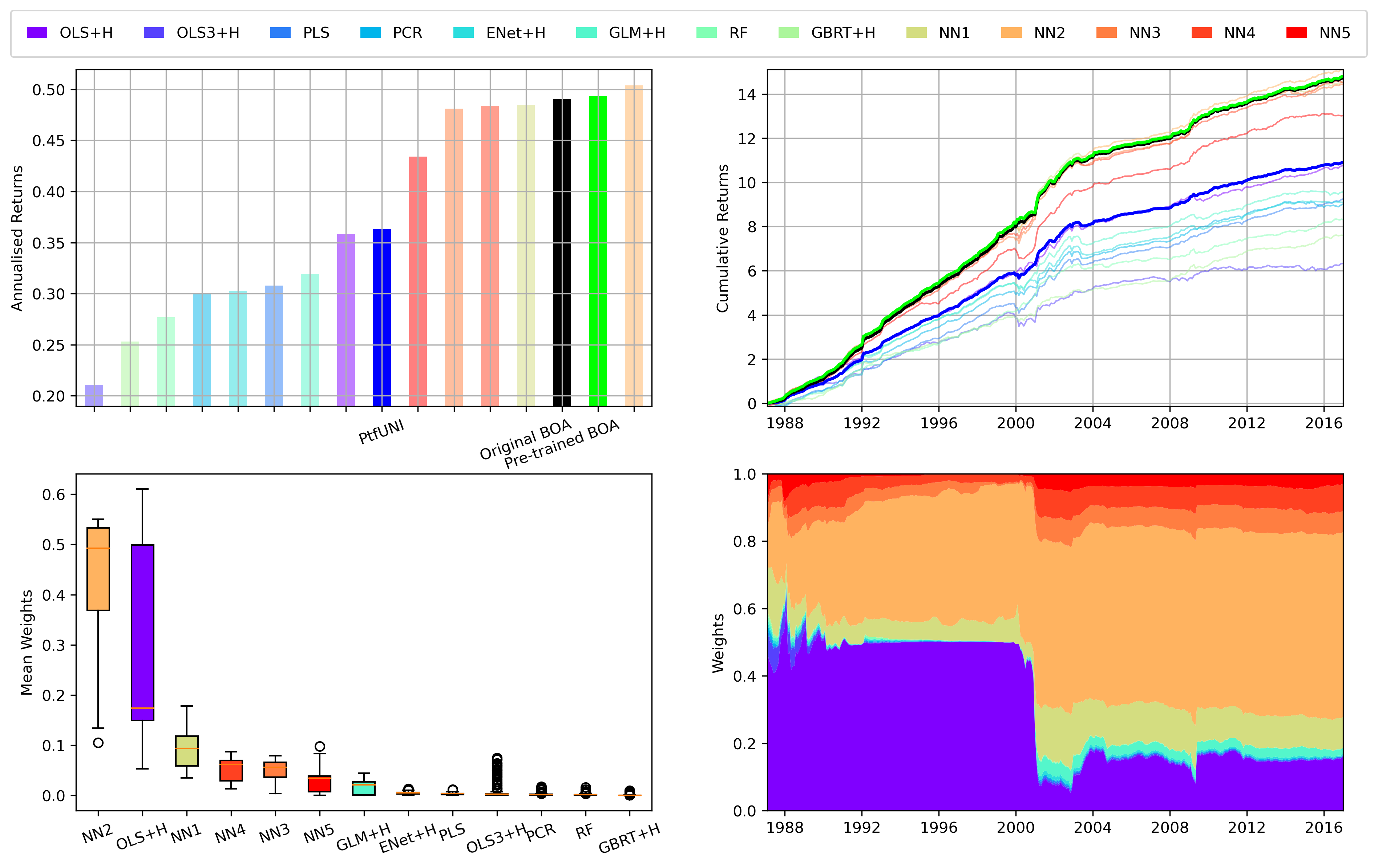}
    \end{center}
    % \vspace{-0.5cm}
    \caption{Pre-trained Aggregation Analysis.}
    Note: \textit{These four graphs present on the first row: average returns (left) and cumulative log returns (right) of the experts, the original PtfBOA, the pre-trained PtfBOA and PtfUNI ; on the second row: average weights of the pre-trained PtfBOA (left) and their evolution on the test period (right). PtfBOA is pre-trained during the year 1986 and then tested on 1987-2016 period.Portfolio are equally weighted.}
    \label{fig:validagg}
\end{figure}

Figure \ref{fig:validagg} presents the portfolio returns and corresponding weights of the pre-trained BOA. The top left and top right graphs display respectively the average monthly returns and the cumulative log returns of all the strategies. 
Like the original PtfBOA, the pre-trained PtfBOA offers on average a slightly lower annual return than the best expert NN2 (49.0\% vs 50.4\%), but a slightly higher return than the original PtfBOA (49.3\%).
The two BOA aggregations have a significantly higher annual return than the naive uniform mixture (36.0\%) and non-neural network experts, all below 48.0\%. Looking at cumulative log returns, PtfBOA also appears more resilient in crisis periods such as 2001.
The bottom left and right graphs display the boxplot of the weights of the individual experts within PtfBOA (left) and their dynamic evolution (right). 
The PtfBOA starts by giving more importance to OLS+H and NN2 and converges faster to the first stationary regime. Similar to the standard aggregation presented in the previous section, OLS+H and neural networks dominate the mixture. 
Table \ref{tab:validagg} reports performance metrics of the best expert, the uniform mixture and the two PtfBOA aggregations. Small improvements come from the validation set prior. The SR of the pre-trained aggregation reaches 2.78, slightly higher than the original aggregation at 2.77. Its volatility is slightly lower (0.17 vs 0.18) and its positive skewness reaches 3.15 (3.11 for the original PtfBOA).

All in all, pre-training enables to retrieve sooner stable weights that converge to the regimes observed in the previous section. Adding some prior information is beneficial for the mixture, which (slightly) improves its portfolio performances compared to the standard aggregation.

\FloatBarrier
%%%%%%%%%%%%%%%%%%%%%%%%%%%%%%%%%%%%%%%%%%%%%%%%%%%%%%%%%%
\subsubsection{Expert Specialisation}
\label{sec:expertspecialisation}

\begin{figure}[h]
    \begin{center}
    \includegraphics[width=\textwidth]{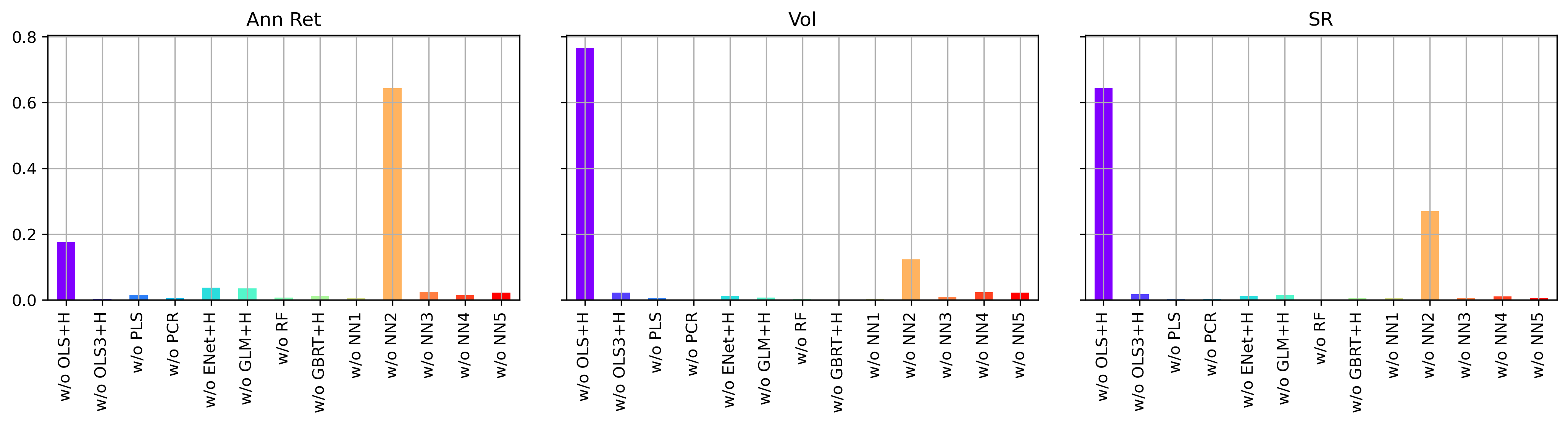}
    \end{center}
    % \vspace{-0.5cm}
    \caption{Experts' Importance in the Bernstein Online Aggregation.}
    Note: \textit{The three graphs display experts importance computed over the out-of-sample period 1987-2016 based on three performance indicators: (1) annualised returns (Ann Ret), (2) volatility (Vol) and (3) annual Sharpe Ratio (SR).
    The importance indicator is obtained by measuring the difference between the performance of the mixture BOA considering all experts and the performance of the mixture when dropping a given expert, and then normalised to sum to one. Portfolios are equally weighted.}
    \label{fig:aggimportance}
    % \vspace{-0.5cm}
\end{figure}

\paragraph*{Expert importance}
To analyse the sensitivity of the results to the set of experts considered in the aggregation, the variation of several portfolio performance metrics are calculated by individually dropping each expert from the mixture.
The study focuses on three performance metrics: annualised average return, volatility and annual Sharpe Ratio over the test period. 
The expert importance is defined as the difference between the performance of the mixture considering all experts and the performance obtained by excluding a given expert. 
The experts' importance indicator is then derived by normalising the differences of all the individual experts to sum to one. 
The expert importance for each performance indicator is given in Figure \ref{fig:aggimportance}.

The aggregation has a lower performance when dropping OLS+H or NN2 and this loss cannot be compensated by other experts or any convex combination of them. Dropping OLS+H from the aggregation leads to a much larger volatility and a smaller SR. OLS+H seems to be a stable expert on which the mixture should rely on average, a somewhat unexpected result given that this linear model can be sensitive to over-fitting in high dimension. On the opposite, the importance of NN2 relies on its high return, at the cost of higher volatility.
These results lead to search how variations of OLS+H or NN2 predictions could affect the mixture. 

\paragraph*{Expert specialisation} An individual expert can provide higher average returns on the test period than the aggregated Portfolio. 
This is an opportunity to improve the overall mixture by adding a slight variation of this best expert to the initial set of experts. To do so, if an expert obtains a lower loss than the aggregation, the expert is split into several new experts by re-calibrating it several times with new parameters or less inputs, as done in \cite{devaine2013forecasting} for time series forecasting.
This new set of experts is added to the initial set and a new aggregation is performed. Several methods to create new expert have been explored in the literature, such as Bagging, Specialisation, Temp Double Scale or Boosting \citep{GaillardGoude2014}.

Based on the analysis of experts' importance, the neural network NN2 and OLS+H are split in several additional experts with the Bagging method, which gives better empirical results and is more computationally efficient.
The method consists in training a bunch of identical models in parallel, where each model is trained by a random subset of the data.
$K^{'}=10$ new Ordinary Least Squared with Huber loss are trained as described in Section \ref{sec:empirical_gu_descrip}, and differ from the original OLS+H in the way they are fed during the training process (containing 80\% of the original data). Models optimised with the Huber loss (noted ``+H'') are trained by gradient descent which could be sensitive to local minima. Bagging allows to make the estimation more robust. Then the new strategies are added to the initial set of experts. Bagging with NN2 is done in a similar way.

\begin{figure}[H]
    \begin{center}
    \includegraphics[width=\textwidth]{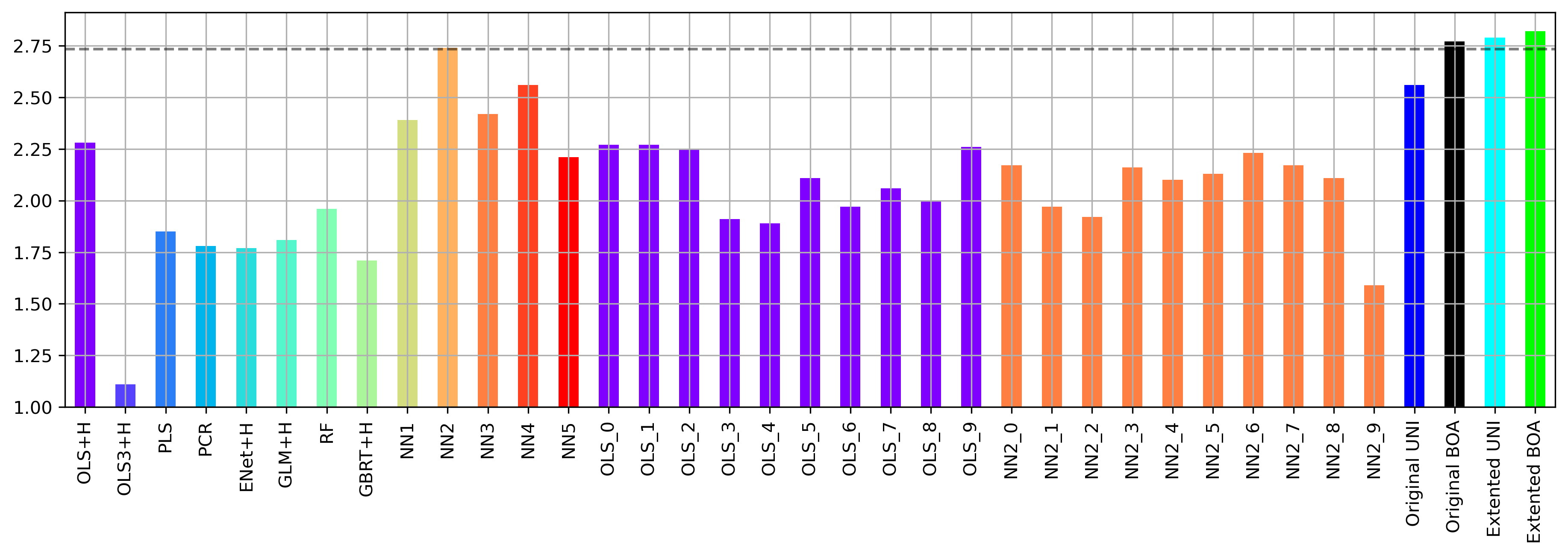}
    \end{center}
    \vspace{-0.5cm}
    \caption{Portfolio Sharpe Ratios with Expert Specialisation.}
    Note: \textit{The graph plots the annual Sharpe Ratios of individual experts and aggregated portfolios, without (Original PtfUNI and PtfBOA) and with (Extended PtfUNI and PtfBOA) expert specialisation. Aggregation is performed with the initial set of $K$ experts plus $K^{'}$ new specialised neural networks NN2 and $K^{'}$ new specialised OLS+H. New forecasting models are trained by Bagging, then the corresponding portfolios are added in the initial set of experts.
    The dash line indicates the best expert Sharpe ratio. Portfolios are equally weighted.
    }
    \label{fig:speexperts}
\end{figure}

Figure \ref{fig:speexperts} shows annual Sharpe Ratios of the specialised aggregation with $K+2K^{'}$ (here 33) experts.
% ($K^{'}$ for OSL+H and $K^{'}$ for NN2). 
The extended PtfBOA outperforms all experts and reaches an annual Sharpe Ratio of 2.82, followed by the extended uniform mixture (2.79) and the original PtfBOA (2.77). More statistics can be found in Table \ref{tab:spedesc} in Appendix, in particular the extended PtfBOA keeps a low maximum monthly loss (7\%). 
Expert specialisation brings Sharpe Ratio improvement while keeping the attractive properties (especially in terms of risk) of the original mixture.

\FloatBarrier

%%%%%%%%%%%%%%%%%%%%%%%%%%%%%%%%%%%%%%%%%%%%%%%%%%%%
%\newpage
\section*{Conclusion}

A portfolio construction methodology based on a sequential aggregation of experts is presented. The strategies, called experts, lie on several forecasting algorithms such as linear models, tree-based models and neural networks. 
The aggregation performs online a convex combination of experts and adapts their weights dynamically according to their performance. 
The originality of the approach is to apply online aggregation directly on strategies, which is an easy way to improve portfolio allocation by combining heterogeneous strategies in a single algorithm.
Online aggregation is particularly promising in finance where market conditions are known to be non-stationary. The aggregation is not computationally costly and considers directly the forecasts of the experts without any assumption on the data distribution and the expert models, allowing to consider any (potentially black-box) strategy. Moreover, the aggregation rules can easily be interpreted and are theoretically grounded. By building long-short strategy based on US stocks, numerical tests show that BOA aggregation offers higher performance than individual experts and simple mixtures. 
Betting on a single expert could be more attractive when one focuses on cumulative returns, but the aggregation oappears to be more robust over time and reduces the risk by deceasing significantly the maximum monthly loss and the maximum draw down.
Further works could examine the design of specific loss functions for portfolio construction, leveraging the easy adaptation of the aggregation framework to alternative objectives.

\section*{Declarations}
This work is supported by  FiME, Laboratoire de Finance des March\'es de l'\'Energie, and EDF Lab.

\medbreak
\noindent\textbf{Conflict of interest} The authors declare that they have no conflict of interest

%%%%%%%%%%%%%%%%%%%%%%%%%%%%%%%%%%%%%%%%%%%%%%%%%%%%
\bibliography{bibliography}

\begin{thebibliography}{57}
\providecommand{\natexlab}[1]{#1}
\providecommand{\url}[1]{\texttt{#1}}
\expandafter\ifx\csname urlstyle\endcsname\relax
  \providecommand{\doi}[1]{doi: #1}\else
  \providecommand{\doi}{doi: \begingroup \urlstyle{rm}\Url}\fi

\bibitem[Abadi et~al.(2015)Abadi, Agarwal, Barham, Brevdo, Chen, Citro,
  Corrado, Davis, Dean, Devin, Ghemawat, Goodfellow, Harp, Irving, Isard, Jia,
  Jozefowicz, Kaiser, Kudlur, Levenberg, Man\'{e}, Monga, Moore, Murray, Olah,
  Schuster, Shlens, Steiner, Sutskever, Talwar, Tucker, Vanhoucke, Vasudevan,
  Vi\'{e}gas, Vinyals, Warden, Wattenberg, Wicke, Yu, and
  Zheng]{tensorflow2015-whitepaper}
M.~Abadi, A.~Agarwal, P.~Barham, E.~Brevdo, Z.~Chen, C.~Citro, G.~S. Corrado,
  A.~Davis, J.~Dean, M.~Devin, S.~Ghemawat, I.~Goodfellow, A.~Harp, G.~Irving,
  M.~Isard, Y.~Jia, R.~Jozefowicz, L.~Kaiser, M.~Kudlur, J.~Levenberg,
  D.~Man\'{e}, R.~Monga, S.~Moore, D.~Murray, C.~Olah, M.~Schuster, J.~Shlens,
  B.~Steiner, I.~Sutskever, K.~Talwar, P.~Tucker, V.~Vanhoucke, V.~Vasudevan,
  F.~Vi\'{e}gas, O.~Vinyals, P.~Warden, M.~Wattenberg, M.~Wicke, Y.~Yu, and
  X.~Zheng.
\newblock {TensorFlow}: Large-scale machine learning on heterogeneous systems,
  2015.
\newblock URL \url{https://www.tensorflow.org/}.
\newblock Software available from tensorflow.org.

\bibitem[Albuquerque et~al.(2022)Albuquerque, Peng, and
  Silva]{albuquer2022quemaking}
P.~H.~M. Albuquerque, Y.~Peng, and J.~P. F.~d. Silva.
\newblock Making the whole greater than the sum of its parts: A literature
  review of ensemble methods for financial time series forecasting.
\newblock \emph{Journal of Forecasting}, 41\penalty0 (8):\penalty0 1701--1724,
  2022.
\newblock \doi{https://doi.org/10.1002/for.2894}.

\bibitem[Amat et~al.(2018)Amat, Michalski, and Stoltz]{amat2018fundamentals}
C.~Amat, T.~Michalski, and G.~Stoltz.
\newblock Fundamentals and exchange rate forecastability with simple machine
  learning methods.
\newblock \emph{Journal of International Money and Finance}, 88:\penalty0
  1--24, 2018.

\bibitem[Atiya(2020)]{atiya2020does}
A.~F. Atiya.
\newblock Why does forecast combination work so well?
\newblock \emph{International Journal of Forecasting}, 36\penalty0
  (1):\penalty0 197--200, 2020.

\bibitem[Auder et~al.(2016)Auder, Bobbia, Poggi, and
  Portier]{auder2016sequential}
B.~Auder, M.~Bobbia, J.-M. Poggi, and B.~Portier.
\newblock Sequential aggregation of heterogeneous experts for pm10 forecasting.
\newblock \emph{Atmospheric Pollution Research}, 7\penalty0 (6):\penalty0
  1101--1109, 2016.

\bibitem[Azoury and Warmuth(2001)]{azoury2001relative}
K.~S. Azoury and M.~K. Warmuth.
\newblock Relative loss bounds for on-line density estimation with the
  exponential family of distributions.
\newblock \emph{Machine Learning}, 43\penalty0 (3):\penalty0 211--246, 2001.

\bibitem[Bai et~al.(2010)Bai, Sun, Luo, and Zhang]{bai2010forecasting}
Y.~Bai, J.~Sun, J.~Luo, and X.~Zhang.
\newblock Forecasting financial time series with ensemble learning.
\newblock \emph{2010 International Symposium on Intelligent Signal Processing
  and Communication Systems}, pages 1--4, 2010.
\newblock \doi{10.1109/ISPACS.2010.5704751}.

\bibitem[Bates and Granger(1969)]{bates1969combination}
J.~M. Bates and C.~W. Granger.
\newblock The combination of forecasts.
\newblock \emph{Journal of the Operational Research Society}, 20\penalty0
  (4):\penalty0 451--468, 1969.

\bibitem[Blackwell(1956)]{blackwell1956analog}
D.~Blackwell.
\newblock An analog of the minimax theorem for vector payoffs.
\newblock \emph{Pacific Journal of Mathematics}, 6\penalty0 (1):\penalty0 1 --
  8, 1956.

\bibitem[Breiman(1996)]{breiman1996bagging}
L.~Breiman.
\newblock Bagging predictors.
\newblock \emph{Machine Learning}, 24:\penalty0 123--140, 1996.

\bibitem[Breiman(2001)]{breiman2001random}
L.~Breiman.
\newblock Random forests.
\newblock \emph{Machine Learning}, 45:\penalty0 5--32, 2001.

\bibitem[Brown et~al.(2005{\natexlab{a}})Brown, Wyatt, Harris, and
  Yao]{brown2005diversity}
G.~Brown, J.~Wyatt, R.~Harris, and X.~Yao.
\newblock Diversity creation methods: a survey and categorisation.
\newblock \emph{Information Fusion}, 6\penalty0 (1):\penalty0 5--20,
  2005{\natexlab{a}}.

\bibitem[Brown et~al.(2005{\natexlab{b}})Brown, Wyatt, Tino, and
  Bengio]{brown2005managing}
G.~Brown, J.~L. Wyatt, P.~Tino, and Y.~Bengio.
\newblock Managing diversity in regression ensembles.
\newblock \emph{Journal of Machine Learning Research}, 6\penalty0 (9),
  2005{\natexlab{b}}.

\bibitem[Butaru et~al.(2016)Butaru, Chen, Clark, Das, Lo, and
  Siddique]{butaru2016risk}
F.~Butaru, Q.~Chen, B.~Clark, S.~Das, A.~W. Lo, and A.~Siddique.
\newblock Risk and risk management in the credit card industry.
\newblock \emph{Journal of Banking \& Finance}, 72:\penalty0 218--239, 2016.

\bibitem[Cesa-Bianchi and Lugosi(2003)]{cesa2003potential}
N.~Cesa-Bianchi and G.~Lugosi.
\newblock Potential-based algorithms in on-line prediction and game theory.
\newblock \emph{Machine Learning}, 51:\penalty0 239--261, 2003.

\bibitem[Cesa-Bianchi and Lugosi(2006)]{cesa2006prediction}
N.~Cesa-Bianchi and G.~Lugosi.
\newblock \emph{Prediction, Learning, and Games}.
\newblock Cambridge University Press, 2006.

\bibitem[Debry and Mallet(2014)]{debry2014ensemble}
E.~Debry and V.~Mallet.
\newblock Ensemble forecasting with machine learning algorithms for ozone,
  nitrogen dioxide and pm10 on the prev'air platform.
\newblock \emph{Atmospheric Environment}, 91:\penalty0 71--84, 2014.

\bibitem[Devaine et~al.(2013)Devaine, Gaillard, Goude, and
  Stoltz]{devaine2013forecasting}
M.~Devaine, P.~Gaillard, Y.~Goude, and G.~Stoltz.
\newblock Forecasting electricity consumption by aggregating specialized
  experts.
\newblock \emph{Machine Learning}, 90\penalty0 (2):\penalty0 231--260, 2013.

\bibitem[Freund et~al.(1996)Freund, Schapire, et~al.]{freund1996experiments}
Y.~Freund, R.~E. Schapire, et~al.
\newblock Experiments with a new boosting algorithm.
\newblock In \emph{{ICML}}, volume~96, pages 148--156. Citeseer, 1996.

\bibitem[Freund et~al.(1997)Freund, Schapire, Singer, and
  Warmuth]{freund1997using}
Y.~Freund, R.~E. Schapire, Y.~Singer, and M.~K. Warmuth.
\newblock Using and combining predictors that specialize.
\newblock In \emph{Proceedings of The Twenty-ninth Annual ACM Symposium on
  Theory of Computing}, pages 334--343, 1997.

\bibitem[Freyberger et~al.(2020)Freyberger, Neuhierl, and
  Weber]{freyberger2020dissecting}
J.~Freyberger, A.~Neuhierl, and M.~Weber.
\newblock Dissecting characteristics nonparametrically.
\newblock \emph{The Review of Financial Studies}, 33\penalty0 (5):\penalty0
  2326--2377, 2020.

\bibitem[Gaillard and Goude(2014)]{GaillardGoude2014}
P.~Gaillard and Y.~Goude.
\newblock Forecasting electricity consumption by aggregating experts; how to
  design a good set of experts.
\newblock \emph{Lecture Notes in Statistics: Modeling and Stochastic Learning
  for Forecasting in High Dimension}, 06 2014.
\newblock \doi{10.1007/978-3-319-18732-7_6}.

\bibitem[Gaillard et~al.(2014)Gaillard, Stoltz, and van
  Erven]{gaillard2014second}
P.~Gaillard, G.~Stoltz, and T.~van Erven.
\newblock A second-order bound with excess losses.
\newblock In M.~F. Balcan, V.~Feldman, and C.~Szepesvári, editors,
  \emph{Proceedings of The 27th Conference on Learning Theory}, volume~35 of
  \emph{Proceedings of Machine Learning Research}, pages 176--196, Barcelona,
  Spain, 13--15 Jun 2014. PMLR.

\bibitem[Gaillard et~al.(2016)Gaillard, Goude, Plagne, Dubois, Thieurmel,
  Gaillard, Rcpp, Rcpp, and Rdpack]{gaillardpackage}
P.~Gaillard, Y.~Goude, L.~Plagne, T.~Dubois, B.~Thieurmel, M.~P. Gaillard,
  L.~Rcpp, R.~I. Rcpp, and R.~Rdpack.
\newblock Package {OPERA}, 2016.

\bibitem[Gu et~al.(2020)Gu, Kelly, and Xiu]{GuKellyXiu2018}
S.~Gu, B.~Kelly, and D.~Xiu.
\newblock Empirical asset pricing via machine learning.
\newblock \emph{The Review of Financial Studies}, 33\penalty0 (5):\penalty0
  2223--2273, 2020.

\bibitem[Hannan(1957)]{hannan1957approximation}
J.~Hannan.
\newblock Approximation to bayes risk in repeated play.
\newblock \emph{Contributions to the Theory of Games}, 3:\penalty0 97--139,
  1957.

\bibitem[Heaton et~al.(2017)Heaton, Polson, and Witte]{heaton2017deep}
J.~B. Heaton, N.~G. Polson, and J.~H. Witte.
\newblock Deep learning for finance: deep portfolios.
\newblock \emph{Applied Stochastic Models in Business and Industry},
  33\penalty0 (1):\penalty0 3--12, 2017.

\bibitem[Herbster and Warmuth(1998)]{herbster1998tracking}
M.~Herbster and M.~K. Warmuth.
\newblock Tracking the best expert.
\newblock \emph{Machine Learning}, 32\penalty0 (2):\penalty0 151--178, 1998.

\bibitem[Hutchinson et~al.(1994)Hutchinson, Lo, and
  Poggio]{hutchinson1994nonparametric}
J.~M. Hutchinson, A.~W. Lo, and T.~Poggio.
\newblock A nonparametric approach to pricing and hedging derivative securities
  via learning networks.
\newblock \emph{The Journal of Finance}, 49\penalty0 (3):\penalty0 851--889,
  1994.

\bibitem[Khandani et~al.(2010)Khandani, Kim, and Lo]{khandani2010consumer}
A.~E. Khandani, A.~J. Kim, and A.~W. Lo.
\newblock Consumer credit-risk models via machine-learning algorithms.
\newblock \emph{Journal of Banking \& Finance}, 34\penalty0 (11):\penalty0
  2767--2787, 2010.

\bibitem[Kozak et~al.(2020)Kozak, Nagel, and Santosh]{kozak2020shrinking}
S.~Kozak, S.~Nagel, and S.~Santosh.
\newblock Shrinking the cross-section.
\newblock \emph{Journal of Financial Economics}, 135\penalty0 (2):\penalty0
  271--292, 2020.

\bibitem[Lewellen(2014)]{lewellen2014cross}
J.~Lewellen.
\newblock The cross section of expected stock returns.
\newblock \emph{Critical Finance Review}, 4\penalty0 (1):\penalty0 1--44, 2014.

\bibitem[Lin et~al.(2021)Lin, Zhou, Liu, and Bian]{lin2021learning}
H.~Lin, D.~Zhou, W.~Liu, and J.~Bian.
\newblock Learning multiple stock trading patterns with temporal routing
  adaptor and optimal transport.
\newblock In \emph{Proceedings of the 27th ACM SIGKDD Conference on Knowledge
  Discovery \& Data Mining}, pages 1017--1026, 2021.

\bibitem[Littlestone and Warmuth(1994)]{littlestone1994weighted}
N.~Littlestone and M.~K. Warmuth.
\newblock The weighted majority algorithm.
\newblock \emph{Information and Computation}, 108\penalty0 (2):\penalty0
  212--261, 1994.

\bibitem[Moritz and Zimmermann(2016)]{MoritzZimmermann2016}
B.~Moritz and T.~Zimmermann.
\newblock Tree-based conditional portfolio sorts: The relation between past and
  future stock returns.
\newblock \emph{Available at SSRN 2740751}, 2016.

\bibitem[Mourtada and Maillard(2017)]{mourtada2017efficient}
J.~Mourtada and O.-A. Maillard.
\newblock Efficient tracking of a growing number of experts.
\newblock In \emph{International Conference on Algorithmic Learning Theory},
  pages 517--539. PMLR, 2017.

\bibitem[Nowotarski and Weron(2018)]{nowotarski2018recent}
J.~Nowotarski and R.~Weron.
\newblock Recent advances in electricity price forecasting: A review of
  probabilistic forecasting.
\newblock \emph{Renewable and Sustainable Energy Reviews}, 81:\penalty0
  1548--1568, 2018.

\bibitem[Nti et~al.(2020)Nti, Adekoya, and Weyori]{nti2020comprehensive}
I.~K. Nti, A.~F. Adekoya, and B.~A. Weyori.
\newblock A comprehensive evaluation of ensemble learning for stock-market
  prediction.
\newblock \emph{Journal of Big Data}, 7\penalty0 (1):\penalty0 1--40, 2020.

\bibitem[Pedregosa et~al.(2011)Pedregosa, Varoquaux, Gramfort, Michel, Thirion,
  Grisel, Blondel, Prettenhofer, Weiss, Dubourg, Vanderplas, Passos,
  Cournapeau, Brucher, Perrot, and Duchesnay]{sklearn}
F.~Pedregosa, G.~Varoquaux, A.~Gramfort, V.~Michel, B.~Thirion, O.~Grisel,
  M.~Blondel, P.~Prettenhofer, R.~Weiss, V.~Dubourg, J.~Vanderplas, A.~Passos,
  D.~Cournapeau, M.~Brucher, M.~Perrot, and E.~Duchesnay.
\newblock Scikit-learn: Machine learning in {P}ython.
\newblock \emph{Journal of Machine Learning Research}, 12:\penalty0 2825--2830,
  2011.

\bibitem[Petropoulos et~al.(2022)Petropoulos, Apiletti, Assimakopoulos, Babai,
  Barrow, Taieb, Bergmeir, Bessa, Bijak, Boylan,
  et~al.]{petropoulos2020forecasting}
F.~Petropoulos, D.~Apiletti, V.~Assimakopoulos, M.~Z. Babai, D.~K. Barrow,
  S.~B. Taieb, C.~Bergmeir, R.~J. Bessa, J.~Bijak, J.~E. Boylan, et~al.
\newblock Forecasting: theory and practice.
\newblock \emph{International Journal of Forecasting}, 2022.

\bibitem[Rapach et~al.(2013)Rapach, Strauss, and Zhou]{rapach2013international}
D.~E. Rapach, J.~K. Strauss, and G.~Zhou.
\newblock International stock return predictability: what is the role of the
  united states?
\newblock \emph{The Journal of Finance}, 68\penalty0 (4):\penalty0 1633--1662,
  2013.

\bibitem[Rasekhschaffe and Jones(2019)]{rasekhschaffe2019machine}
K.~C. Rasekhschaffe and R.~C. Jones.
\newblock Machine learning for stock selection.
\newblock \emph{Financial Analysts Journal}, 75\penalty0 (3):\penalty0 70--88,
  2019.

\bibitem[Sadhwani et~al.(2021)Sadhwani, Giesecke, and
  Sirignano]{sadhwani2021deep}
A.~Sadhwani, K.~Giesecke, and J.~Sirignano.
\newblock Deep learning for mortgage risk.
\newblock \emph{Journal of Financial Econometrics}, 19\penalty0 (2):\penalty0
  313--368, 2021.

\bibitem[Schapire(1990)]{schapire1990strength}
R.~E. Schapire.
\newblock The strength of weak learnability.
\newblock \emph{Machine Learning}, 5:\penalty0 197--227, 1990.

\bibitem[Stoltz(2005)]{Stoltz2005}
G.~Stoltz.
\newblock \emph{{Incomplete Information and Internal Regret in Prediction of
  Individual Sequences}}.
\newblock PhD thesis, {Universit{\'e} Paris Sud - Paris XI}, 2005.
\newblock URL \url{https://tel.archives-ouvertes.fr/tel-00009759}.

\bibitem[Sun et~al.(2018)Sun, Wei, and Wang]{sun2018adaboost}
S.~Sun, Y.~Wei, and S.~Wang.
\newblock Adaboost-lstm ensemble learning for financial time series
  forecasting.
\newblock In \emph{Computational Science--ICCS 2018: 18th International
  Conference, Wuxi, China, June 11--13, 2018 Proceedings, Part III 18}, pages
  590--597. Springer, 2018.

\bibitem[Taillardat et~al.(2016)Taillardat, Mestre, Zamo, and
  Naveau]{taillardat2016calibrated}
M.~Taillardat, O.~Mestre, M.~Zamo, and P.~Naveau.
\newblock Calibrated ensemble forecasts using quantile regression forests and
  ensemble model output statistics.
\newblock \emph{Monthly Weather Review}, 144\penalty0 (6):\penalty0 2375--2393,
  2016.

\bibitem[Thorey et~al.(2017)Thorey, Mallet, and Baudin]{thorey2017online}
J.~Thorey, V.~Mallet, and P.~Baudin.
\newblock Online learning with the continuous ranked probability score for
  ensemble forecasting.
\newblock \emph{Quarterly Journal of the Royal Meteorological Society},
  143\penalty0 (702):\penalty0 521--529, 2017.

\bibitem[Vovk(1997)]{vovk1997competitive}
V.~Vovk.
\newblock Competitive on-line linear regression.
\newblock \emph{Advances in Neural Information Processing Systems}, 10, 1997.

\bibitem[Vovk(1998)]{vovk1998game}
V.~Vovk.
\newblock A game of prediction with expert advice.
\newblock \emph{Journal of Computer and System Sciences}, 56\penalty0
  (2):\penalty0 153--173, 1998.
\newblock ISSN 0022-0000.
\newblock \doi{https://doi.org/10.1006/jcss.1997.1556}.

\bibitem[Vovk(2006)]{vovk2006line}
V.~Vovk.
\newblock On-line regression competitive with reproducing kernel hilbert
  spaces.
\newblock In \emph{International Conference on Theory and Applications of
  Models of Computation}, pages 452--463. Springer, 2006.

\bibitem[Vovk(1990)]{vovk1990aggregating}
V.~G. Vovk.
\newblock Aggregating strategies.
\newblock \emph{Proc. of Computational Learning Theory, 1990}, page 371–386,
  1990.

\bibitem[Weng et~al.(2018)Weng, Lu, Wang, Megahed, and
  Martinez]{weng2018predicting}
B.~Weng, L.~Lu, X.~Wang, F.~M. Megahed, and W.~Martinez.
\newblock Predicting short-term stock prices using ensemble methods and online
  data sources.
\newblock \emph{Expert Systems with Applications}, 112:\penalty0 258--273,
  2018.

\bibitem[Wintenberger(2017)]{wintenberger2017optimal}
O.~Wintenberger.
\newblock Optimal learning with bernstein online aggregation.
\newblock \emph{Machine Learning}, 106\penalty0 (1):\penalty0 119--141, 2017.

\bibitem[WRDS()]{WRDSdatabase}
WRDS.
\newblock Wharton research data services.
\newblock URL \url{https://wrds-web.wharton.upenn.edu/wrds/}.

\bibitem[Yang et~al.(2020)Yang, Liu, Zhong, and Walid]{yang2020deep}
H.~Yang, X.-Y. Liu, S.~Zhong, and A.~Walid.
\newblock Deep reinforcement learning for automated stock trading: An ensemble
  strategy.
\newblock In \emph{Proceedings of the first ACM International Conference on AI
  in Finance}, pages 1--8, 2020.

\bibitem[Yao et~al.(2000)Yao, Li, and Tan]{yao2000option}
J.~Yao, Y.~Li, and C.~L. Tan.
\newblock Option price forecasting using neural networks.
\newblock \emph{Omega}, 28\penalty0 (4):\penalty0 455--466, 2000.

\end{thebibliography}

%%%%%%%%%%%%%%%%%%%%%%%%%%%%%%%%%%%%%%%%%%%%%%%%%%%%
\newpage
\appendix
\section*{Appendix}\label{sec:annex}

\section{Algorithms and Variables}\label{sec:algorithms}

\renewcommand{\thetable}{\Alph{section}.\arabic{table}}
\renewcommand{\thefigure}{\Alph{section}.\arabic{figure}}

In this section, some details are provided about the data and the parameters of the experts.
% For more elements about theoretical models or algorithms, see appendix of \cite{GuKellyXiu2018}.
All the variables used in this paper are exactly the same as \cite{GuKellyXiu2018} and precisely described in the Table A.6 of its Appendix.

\begin{table}[H]
\begin{center}
\resizebox{\textwidth}{!}{
\begin{tabular}{l cccc}
\toprule
           &OLS-OLS3+H                     &PLS                         &PCR&ENet+H\\
\midrule
Huber Loss &$\xi=0.999$&-&-&$\xi=0.999$ \\
Hyper Param& &P=94 &P=94& $\alpha\in(10e^{-4},10e^{-1})$ \\
           & Ensemble: 10 & & &$\rho=0.5$ \\
           &&&&Ensemble: 10\\    
\bottomrule
\toprule
    &GLM+H&RF&GBRT+H&NN1-NN5\\
\midrule
Huber Loss &$\xi=0.999$&-&$\xi=0.999$        &-               \\
Hyper Param&$\alpha\in(10e^{-4},10e^{-1})$& Nb trees: 300  & Nb trees: 1000       &Batch size: 10000\\
           &Ensemble: 10& Depth$\in(1,6)$& Depth$\in(1,2)$     &Nb epoch: 100\\
           && Bootstrap:True & Learning rate: \{0.01,0.1\} &Learning rate: 0.01\\
           &&                &                     &Adam: Default \\
           &&                &                     &Ensemble: 10\\
           &&                &                     &Patience: 10\\
           &&                &                     &L1 pen$\in(10^{-5},10^{-3})$\\
\bottomrule
\end{tabular}
}
\end{center}
\caption{Description of Hyper-parameters of the Forecasting Models}
Note: \textit{P=94 number of variables. OLS3+H only includes variables mom12m, size, bm. Hyper-parameters are optimised with the validation set.}
\label{tab:hyperparameters}
\end{table}
Neural networks denoted NN1, NN2, NN3, NN4 and NN5 have hidden layer(s) of 32, (32,16), (32,16,8), (32,16,8,4) and (32,16,8,4,2) nodes respectively. ReLU activation function is used for each hidden layer, and regularisation methods include batch normalisation\footnote{The standard scale is used: $(x - \mu)/ \sigma$, with $\mu, \sigma$ mean and standard deviation respectively.},
% \cite{batchnorm}
learning rate shrinking (Adam),
% \cite{adam}
early stopping 
and ensemble. These setups are common standards in deep learning literature.\\

\begin{table}[H]
\begin{center}
\begin{tabular}{l ccc  c cc}
\toprule
      &OLS3&OLS7&OLS15&&RF&NN3\\
\midrule
\%R2&0.16&0.19&0.19&&0.19&0.45\\
% S&P500&-0.20&0.27&0.27&0.25\\
SR&0.95&1.21&1.33&&1.96&2.42\\
Turnover&0.49&0.48&0.56&&0.92&1.20\\
\bottomrule
\end{tabular}
\end{center}
\caption{OLS Benchmark Models.}
Note: \textit{This table reports the performance of different Ordinary Least Squared benchmark models, as well as RF and NN3. The predictive R2 for stock return forecasting on test period, the Sharpe ratios of long-short strategies as well as their turnover are reported. OLS3 includes variables mom12m, size, bm, OLS7 adds acc, roaq, agr, egr, and OLS15 adds dy, mom36m, beta, retvol, turn, lev, sp.  Models are trained on training and validation set data, as there is no hyper-parameters. This benchmark can be compared with Table A.11 of \cite{GuKellyXiu2018}.}
\label{tab:benchmark}

\

The forecasting models are implemented in Python using scikit-learn \cite{sklearn} and tensorflow \cite{tensorflow2015-whitepaper} packages.
The online expert aggregation algorithm is computed with the R package OPERA \citep{gaillardpackage}.
\end{table}

\section{Metric Definitions}\label{sec:metrics}

Metrics used in this paper are detailed here. 
Let be $y_{t}$ and $\hat{y}_{t}$ respectively the observed and the prediction values of one asset. 
Mean squared error (MSE) is defined by
$$\ell_{MSE}(y_t,\hat{y}_t) = \frac{1}{T}\sum^T_{t=1} (y_{t} - \hat{y}_{t})^2,$$
and the Huber loss (HL) by
\begin{eqnarray}\ell_{HL}(y_t,\hat{y}_t; \xi) = \frac{1}{T}\sum^T_{t=1} H(y_{t} - \hat{y}_{t}; \xi) \mbox{, where  } H(x;\xi)=
\begin{cases}
x^2 &\mbox{if } |x|\leq\xi\\
2\epsilon|x|- \xi^2 &\mbox{if } |x| \geq \xi.\\
\end{cases}\nonumber
\label{eq:huber}
\end{eqnarray}
The Huber loss  is less sensitive to outliers in the data distribution than MSE. $\xi$ determines the threshold from which it is less important to make an error. 
MSE and HL are used to train the forecasting models, aggregation optimises MSE.
The R2 is used to evaluate the accuracy of asset return estimation: $$R2 = 1- \frac{\sum^T_{t=1} (y_t - \hat{y_t})^2}{\sum^T_{t=1} y_t^2}.$$

\noindent The maximum draw down, 1-month maximum loss and portfolio turnover are defined by: 
\begin{eqnarray}
% \hbox{Ann. return} &=& \frac{12}{T}\sum_{t=1}^T r_t\nonumber\\\nonumber
% \hbox{Ann. vol} &=& \frac{12}{T}\sum_{t=1}^T (r_t - \frac{1}{T}\sum_{t=1}^T r_t)^2\\\nonumber
% \hbox{Skew} &=& \frac{1}{T}\sum_{t=1}^T\frac{ (r_t - \frac{1}{T}\sum_{t=1}^T r_t)^3}{\sqrt{\frac{1}{T}\sum_{t=1}^T (r_t - \frac{1}{T}\sum_{t=1}^T r_t)^2}}\\\nonumber
% \hbox{SR} &=& \sqrt{12} \hbox{Ann return} / \hbox{Vol}\\\nonumber
\hbox{Max DD:} && \max_{0\leq t_1 \leq t_2 \leq T} (cr_{t_1} - cr_{t_2})\nonumber\\
\hbox{Max Loss:}&&  - \min_{0\leq t \leq T} r_t\nonumber\\
\hbox{Turnover:}&& \frac{1}{T}\sum_{t=1}^T\left(\sum_i \left|w_{i,t+1} - w_{i,t}(1+r_{i,t+1})\right|\right)\nonumber
\end{eqnarray}
where $cr_t, r_t$ are the cumulative log return and the monthly excess return of a strategy at $t$ respectively, and $T$ the number of dates in the test period.

\section{Additional Tables and Figures}\label{sec:stockreturns}

\renewcommand{\thetable}{\Alph{section}.\arabic{table}}
\renewcommand{\thefigure}{\Alph{section}.\arabic{figure}}

\begin{figure}[H]
    \begin{center}
    \includegraphics[width=\textwidth]{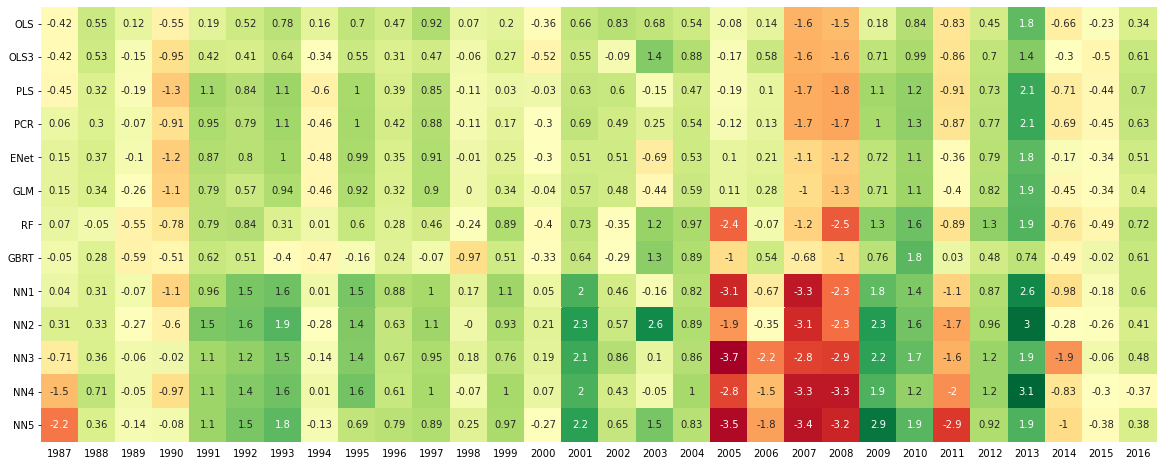}
    \end{center}
    % \vspace{-0.5cm}
    \caption{\%R2 Scores by Year of the Forecasting Models.}
    Note: \textit{The heatmap reports the percent of R2 scores (1\%R2 = 0.01 R2) by year of the thirteen forecasting models OLS+H, OLS3+H, PLS, PCR, ENet+H, GLM+H, RF, GBRT+H, NN1, NN2, NN3, NN4, and NN5 on the out-of-the-sample test period 1987-2016.}
    \label{fig:r2heatmap}
\end{figure}

Performance of the forecasting models are reported in Table \ref{tab:r2scores} and Figure \ref{fig:r2heatmap}. The R2 scores for predicting stock market returns show that the advantage of neural networks over linear models is not decisive. Moreover, while their accuracy is clearly convincing in some years, neural networks and random forest suffer from significant forecasting errors in other periods. 
These unstable scores complicate the choice of the most appropriate algorithm and light out why aggregation is appealing.
Besides, on average the lowest market caps are better estimated than the top market cap stocks. 

\begin{figure}[H]
    \begin{center}
    \includegraphics[width=\textwidth]{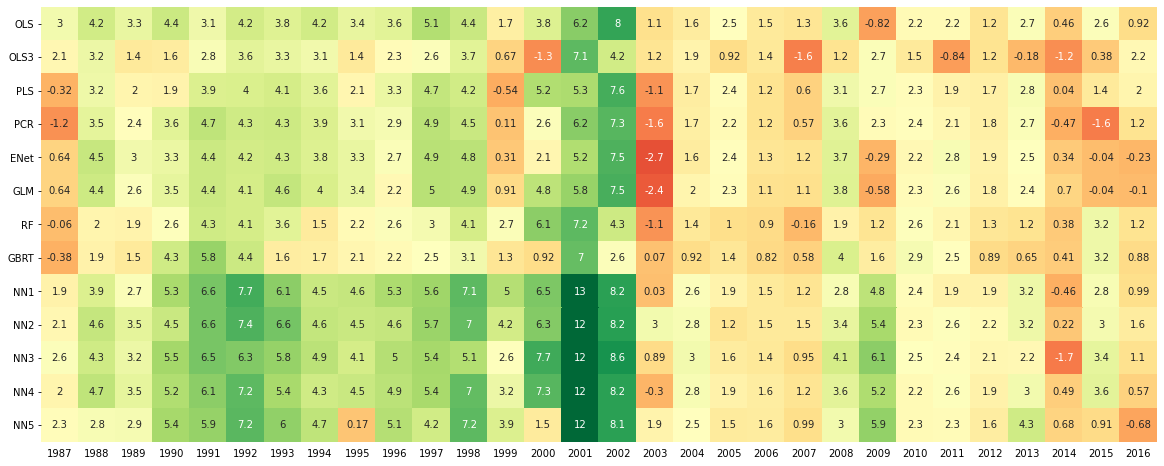}\\
    \includegraphics[width=\textwidth]{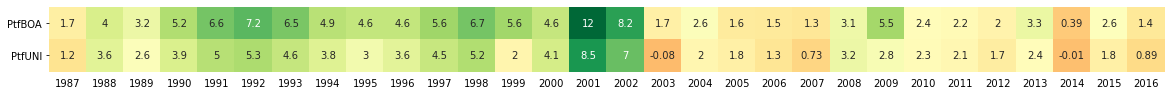}
    \end{center}
    % \vspace{-0.5cm}
    \caption{Average Annualised Returns per Year of Portfolios.}
    Note: \textit{The expert portfolios are equally weighted and computed on the 1987-2016 test period. Experts include OLS+H, OLS3+H, PLS, PCR, ENet+H, GLM+H, RF, GBRT+H and the five neural networks (NN1-NN5). 
    % "+H" indicates the use of Huber loss for the training process.
    PtfBOA is the portfolio obtained with the Bernstein Online Aggregation and PtfUNI is an uniform mixture of the experts.}
    \label{fig:aretheatmap}
\end{figure}

Figure \ref{fig:aretheatmap} illustrates annualised average returns of each expert's long short strategy from 1987 to 2016. Up to 2002, strategies are profitable. However in 2003, several experts suffer from a breakout and are not able to retrieve the same performance afterward, even if models are re-calibrated each year. 
The variation in the rankings of the best experts from year to year emphasises the usefulness of aggregation techniques.

\begin{table}[H]
\begin{center}
\includegraphics[width=\textwidth]{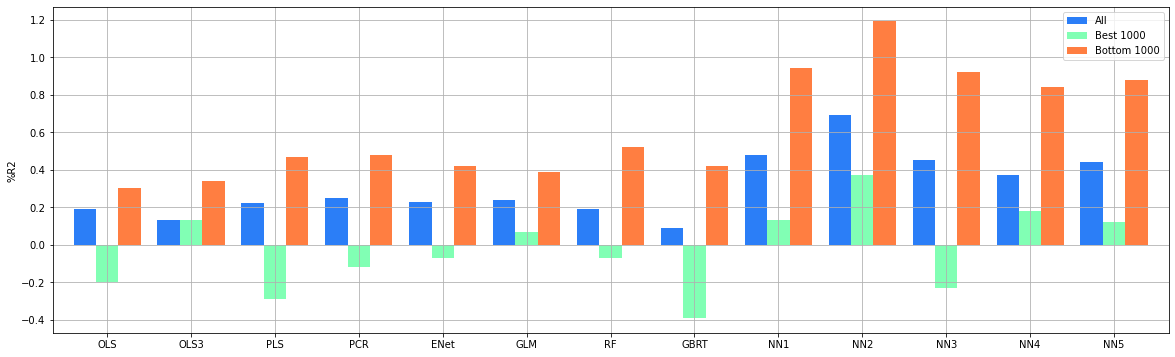}
\resizebox{\textwidth}{!}{
\begin{tabular}{c cccc cccc ccccc}
\toprule
\%R2    &OLS&OLS3&PLS&PCR&ENet&GLM&RF&GBRT&NN1&NN2&NN3&NN4&NN5\\
        &+H&+H&&&+H&+H&&+H&&&&&\\
\midrule
All&0.19&0.13&0.22&0.25&0.23&0.24&0.19&0.09&0.48&0.69&0.45&0.37&0.44\\
Top&-0.20&0.13&-0.29&-0.12&-0.07&0.07&-0.07&-0.39&0.13&0.37&-0.23&0.18&0.12\\
Bot&0.30&0.34&0.47&0.48&0.42&0.39&0.52&0.42&0.94&1.19&0.92&0.84&0.88\\
\bottomrule
\end{tabular}
}
\end{center}
% \vspace{-0.5cm}
\caption{\%R2 Scores of the Forecasting Models}
Note: \textit{The table reports the percent of R2 scores (1\%R2 = 0.01 R2) of the thirteen forecasting models 
OLS+H, OLS3+H, PLS, PCR, ENet+H, GLM+H, RF, GBRT+H, NN1, NN2, NN3, NN4, and NN5 
on the out-of-the-sample test period 1987-2016.
All indicates all the US market universe. Top (resp. Bot) is the top 1000 (resp. bottom 1000) market capitalisation assets.
}
\label{tab:r2scores}
\end{table}

\begin{table}
\begin{center}
\resizebox{0.83\textwidth}{!}{
\begin{tabular}{c cccc c cccc c cccc }
\toprule
&\multicolumn{4}{c}{}&&\multicolumn{4}{c}{OLS+H}&&\multicolumn{4}{c}{}\\
Decile&&&&&&Pred&Real&Std&SR &&&&&\\
\midrule
L&&&&&&-1.35&-0.61&6.95&-0.30&&&&&\\
2&&&&&&-0.41&0.51&6.17&0.29&&&&&\\
3&&&&&&-0.05&0.77&5.61&0.48&&&&&\\
4&&&&&&0.39&0.97&5.23&0.64&&&&&\\
5&&&&&&0.70&1.09&5.07&0.74&&&&&\\
6&&&&&&0.98&1.21&5.05&0.84&&&&&\\
7&&&&&&1.28&1.30&5.05&0.89&&&&&\\
8&&&&&&1.61&1.49&5.23&0.98&&&&&\\
9&&&&&&2.04&1.70&5.45&1.08&&&&&\\
H&&&&&&2.88&2.11&6.00&1.22&&&&&\\
H-L&&&&&&4.49&2.99&4.50&2.28&&&&&\\
% \bottomrule
\toprule
&\multicolumn{4}{c}{OLS3+H}&&\multicolumn{4}{c}{PLS}&&\multicolumn{4}{c}{PCR}\\
Decile&Pred&Real&Std&SR &&Pred&Real&Std&SR &&Pred&Real&Std&SR\\
\midrule
L&-0.19&0.44&6.77&0.23&&-1.03&-0.12&7.22&-0.06&&-0.9&-0.12&6.86&-0.06\\
2&0.16&0.72&5.48&0.46&&-0.22&0.55&6.08&0.31&&-0.15&0.58&6.0&0.33\\
3&0.38&0.98&4.83&0.7&&0.2&0.79&5.55&0.49&&0.24&0.81&5.37&0.52\\
4&0.58&1.00&4.49&0.77&&0.52&0.89&5.21&0.59&&0.54&0.9&5.21&0.6\\
5&0.77&0.98&4.43&0.77&&0.82&0.96&5.05&0.66&&0.82&0.99&5.12&0.67\\
6&0.94&0.96&4.79&0.69&&1.10&1.07&5.16&0.72&&1.08&1.03&5.20&0.69\\
7&1.11&1.06&5.31&0.69&&1.40&1.17&5.18&0.79&&1.35&1.18&5.27&0.78\\
8&1.28&1.22&6.09&0.70&&1.74&1.39&5.25&0.92&&1.66&1.42&5.27&0.94\\
9&1.49&1.23&6.44&0.66&&2.17&1.65&5.35&1.07&&2.04&1.64&5.58&1.02\\
H&1.81&1.94&8.08&0.83&&3.00&2.18&5.87&1.29&&2.79&2.11&6.14&1.19\\
H-L&2.26&1.76&5.47&1.11&&4.29&2.57&4.79&1.85&&3.96&2.50&4.85&1.78\\
% \bottomrule
\toprule
&\multicolumn{4}{c}{ENet+H}&&\multicolumn{4}{c}{GLM+H}&&\multicolumn{4}{c}{RF}\\
Decile&Pred&Real&Std&SR &&Pred&Real&Std&SR &&Pred&Real&Std&SR\\
\midrule
L&-0.86&-0.21&7.24&-0.1&&-0.78&-0.29&7.41&-0.13&&0.20&0.37&7.13&0.18\\
2&-0.16&0.61&6.28&0.34&&-0.13&0.64&6.26&0.35&&0.47&0.63&5.87&0.37\\
3&0.2&0.88&5.62&0.54&&0.23&0.87&5.65&0.53&&0.62&0.71&5.73&0.43\\
4&0.49&0.97&5.19&0.65&&0.50&0.95&5.20&0.63&&0.73&0.94&5.51&0.59\\
5&0.75&1.09&5.03&0.75&&0.75&1.09&5.07&0.74&&0.85&1.04&5.59&0.65\\
6&0.99&1.08&4.94&0.76&&0.99&1.12&4.93&0.79&&1.08&1.07&5.02&0.74\\
7&1.25&1.21&5.01&0.84&&1.23&1.17&5.09&0.80&&1.22&1.09&4.7&0.80\\
8&1.52&1.29&5.20&0.86&&1.49&1.25&5.08&0.85&&1.35&1.05&4.84&0.75\\
9&1.87&1.56&5.67&0.95&&1.82&1.64&5.59&1.02&&1.50&1.23&5.03&0.85\\
H&2.51&2.05&6.04&1.18&&2.42&2.11&5.92&1.23&&2.48&2.42&7.25&1.16\\
H-L&3.63&2.53&4.90&1.77&&3.47&2.66&5.06&1.81&&2.54&2.31&4.05&1.96\\
% \bottomrule
\toprule
&\multicolumn{4}{c}{GBRT+H}&&\multicolumn{4}{c}{NN1}&&\multicolumn{4}{c}{NN2}\\
Decile&Pred&Real&Std&SR &&Pred&Real&Std&SR &&Pred&Real&Std&SR\\
\midrule
L&-0.12&0.14&6.91&0.07&&-1.46&-0.59&8.17&-0.25&&-1.26&-0.63&8.07&-0.27\\
2&0.21&0.81&5.73&0.49&&-0.33&0.46&6.34&0.25&&-0.12&0.29&6.26&0.16\\
3&0.36&0.98&5.45&0.62&&0.17&0.70&5.38&0.45&&0.36&0.59&5.28&0.39\\
4&0.49&0.94&5.32&0.61&&0.52&0.81&4.88&0.57&&0.68&0.79&4.83&0.57\\
5&0.62&1.16&5.52&0.73&&0.82&0.91&4.66&0.68&&0.95&0.90&4.64&0.67\\
6&0.80&1.06&4.95&0.74&&1.09&1.05&4.58&0.79&&1.20&1.10&4.54&0.84\\
7&1.05&1.13&4.75&0.82&&1.38&1.18&4.62&0.88&&1.46&1.21&4.64&0.9\\
8&1.24&1.12&4.83&0.8&&1.72&1.32&4.73&0.97&&1.76&1.35&4.70&1.00\\
9&1.44&1.23&5.72&0.75&&2.19&1.52&5.07&1.04&&2.19&1.63&5.14&1.10\\
H&2.26&1.98&7.76&0.88&&4.12&3.18&8.62&1.28&&4.09&3.31&8.61&1.33\\
H-L&2.65&2.11&4.25&1.71&&5.84&4.04&5.80&2.39&&5.61&4.20&5.27&2.74\\
% \bottomrule
\toprule
&\multicolumn{4}{c}{NN3}&&\multicolumn{4}{c}{NN4}&&\multicolumn{4}{c}{NN5}\\
Decile&Pred&Real&Std&SR &&Pred&Real&Std&SR &&Pred&Real&Std&SR\\
\midrule
L&-1.54&-0.58&8.27&-0.24&&-1.49&-0.61&8.36&-0.25&&-0.87&-0.43&8.03&-0.19\\
2&-0.32&0.44&6.41&0.24&&-0.17&0.40&6.38&0.22&&0.16&0.49&6.22&0.27\\
3&0.20&0.68&5.31&0.44&&0.38&0.70&5.35&0.45&&0.55&0.81&5.27&0.53\\
4&0.55&0.75&4.91&0.53&&0.74&0.80&4.84&0.57&&0.82&0.92&4.72&0.68\\
5&0.85&0.88&4.71&0.65&&1.03&0.92&4.61&0.69&&1.04&1.04&4.65&0.78\\
6&1.11&1.07&4.59&0.81&&1.30&1.08&4.54&0.83&&1.24&0.98&4.60&0.74\\
7&1.39&1.19&4.63&0.89&&1.58&1.22&4.55&0.93&&1.45&1.16&4.61&0.87\\
8&1.70&1.38&4.72&1.01&&1.9&1.35&4.74&0.99&&1.71&1.26&4.74&0.92\\
9&2.15&1.58&5.05&1.09&&2.36&1.53&5.10&1.04&&2.09&1.39&5.31&0.90\\
H&4.12&3.16&8.43&1.30&&4.48&3.16&8.33&1.31&&4.10&2.92&8.81&1.15\\
H-L&5.92&4.01&5.70&2.42&&6.23&4.03&5.56&2.58&&5.23&3.62&5.64&2.21\\
\bottomrule
\end{tabular}
}
\caption{Performance of Equally Weighted Portfolios by Decile.}
\label{tab:decile_equally}
\end{center}
\vspace{-0.2cm}
Note: \textit{Performance on decile of each model for equally weighted portfolios. Rows L, H and HL stand for Low, High and High-minus-Low deciles respectively. Columns Pred, Real, Std, and SR are average predicted monthly returns (in \%), average realised monthly returns (in \%), realised monthly standard deviation (in \%) and Sharpe ratio, respectively.}
\end{table}

\begin{table}
\begin{center}
\resizebox{0.83\textwidth}{!}{
\begin{tabular}{c cccc c cccc c cccc }
\toprule
&\multicolumn{4}{c}{}&&\multicolumn{4}{c}{OLS+H}&&\multicolumn{4}{c}{}\\
Decile&&&&&&Pred&Real&Std&SR &&&&&\\
\midrule
L&&&&&&-1.13&-0.69&4.99&-0.48&&&&&\\
2&&&&&&-0.47&0.12&4.40&0.09&&&&&\\
3&&&&&&-0.15&0.31&3.95&0.28&&&&&\\
4&&&&&&0.09&0.46&3.68&0.43&&&&&\\
5&&&&&&0.30&0.54&3.58&0.53&&&&&\\
6&&&&&&0.50&0.63&3.56&0.62&&&&&\\
7&&&&&&0.71&0.70&3.58&0.68&&&&&\\
8&&&&&&0.94&0.84&3.73&0.78&&&&&\\
9&&&&&&1.23&1.00&3.87&0.89&&&&&\\
H&&&&&&1.82&1.29&4.27&1.04&&&&&\\
H-L&&&&&&3.20&2.24&3.34&2.32&&&&&\\
\toprule
&\multicolumn{4}{c}{OLS3+H}&&\multicolumn{4}{c}{PLS}&&\multicolumn{4}{c}{PCR}\\
Decile&Pred&Real&Std&SR &&Pred&Real&Std&SR &&Pred&Real&Std&SR\\
\midrule
L&-0.32&0.07&4.95&0.05&&-0.90&-0.32&5.30&-0.21&&-0.82&-0.34&4.98&-0.23\\
2&-0.07&0.29&3.95&0.25&&-0.33&0.16&4.38&0.13&&-0.29&0.18&4.30&0.14\\
3&0.08&0.47&3.37&0.48&&-0.04&0.34&3.96&0.29&&-0.02&0.35&3.82&0.32\\
4&0.22&0.48&3.11&0.54&&0.19&0.41&3.70&0.38&&0.19&0.41&3.70&0.38\\
5&0.35&0.47&3.08&0.53&&0.39&0.45&3.57&0.44&&0.39&0.48&3.63&0.46\\
6&0.47&0.45&3.37&0.47&&0.59&0.54&3.64&0.51&&0.57&0.51&3.67&0.48\\
7&0.59&0.52&3.79&0.48&&0.80&0.61&3.65&0.58&&0.76&0.62&3.73&0.57\\
8&0.71&0.65&4.39&0.51&&1.04&0.77&3.70&0.72&&0.98&0.79&3.71&0.73\\
9&0.85&0.65&4.64&0.49&&1.33&0.95&3.73&0.88&&1.25&0.94&3.92&0.83\\
H&1.07&1.16&5.77&0.69&&1.91&1.32&4.06&1.13&&1.77&1.29&4.29&1.04\\
H-L&1.66&1.35&4.14&1.13&&3.08&1.91&3.58&1.85&&2.86&1.89&3.54&1.84\\
\toprule
&\multicolumn{4}{c}{ENet+H}&&\multicolumn{4}{c}{GLM+H}&&\multicolumn{4}{c}{RF}\\
Decile&Pred&Real&Std&SR &&Pred&Real&Std&SR &&Pred&Real&Std&SR\\
\midrule
L&-0.79&-0.39&5.23&-0.26&&-0.73&-0.46&5.36&-0.30&&-0.05&0.04&5.08&0.02\\
2&-0.3&0.20&4.47&0.15&&-0.27&0.21&4.48&0.16&&0.14&0.20&4.14&0.17\\
3&-0.04&0.40&3.98&0.35&&-0.03&0.39&4.03&0.33&&0.26&0.26&4.08&0.22\\
4&0.16&0.47&3.68&0.44&&0.17&0.45&3.68&0.43&&0.34&0.44&3.83&0.39\\
5&0.34&0.54&3.54&0.53&&0.34&0.55&3.59&0.53&&0.43&0.52&3.88&0.46\\
6&0.51&0.54&3.49&0.54&&0.51&0.57&3.48&0.56&&0.58&0.54&3.54&0.53\\
7&0.69&0.63&3.53&0.62&&0.68&0.61&3.60&0.58&&0.68&0.54&3.36&0.56\\
8&0.89&0.70&3.68&0.66&&0.86&0.66&3.59&0.64&&0.77&0.52&3.44&0.52\\
9&1.13&0.88&4.05&0.75&&1.09&0.95&3.94&0.83&&0.88&0.65&3.57&0.63\\
H&1.58&1.24&4.3&1.00&&1.51&1.29&4.16&1.07&&1.54&1.52&5.17&1.02\\
H-L&2.63&1.90&3.64&1.81&&2.5&2.01&3.74&1.86&&1.85&1.74&2.93&2.06\\
\toprule
&\multicolumn{4}{c}{GBRT+H}&&\multicolumn{4}{c}{NN1}&&\multicolumn{4}{c}{NN2}\\
Decile&Pred&Real&Std&SR &&Pred&Real&Std&SR &&Pred&Real&Std&SR\\
\midrule
L&-0.26&-0.13&4.81&-0.09&&-1.21&-0.69&6.00&-0.40&&-1.07&-0.70&5.89&-0.41\\
2&-0.03&0.34&3.96&0.29&&-0.42&0.08&4.58&0.06&&-0.27&-0.04&4.52&-0.03\\
3&0.07&0.47&3.86&0.42&&-0.07&0.27&3.83&0.24&&0.07&0.20&3.76&0.18\\
4&0.16&0.44&3.75&0.41&&0.18&0.34&3.43&0.34&&0.29&0.34&3.41&0.34\\
5&0.26&0.60&3.90&0.54&&0.39&0.42&3.25&0.44&&0.48&0.41&3.25&0.44\\
6&0.38&0.52&3.56&0.51&&0.58&0.52&3.20&0.56&&0.65&0.55&3.17&0.61\\
7&0.57&0.57&3.39&0.58&&0.78&0.62&3.23&0.66&&0.84&0.63&3.24&0.68\\
8&0.70&0.57&3.46&0.57&&1.02&0.72&3.29&0.76&&1.05&0.74&3.28&0.78\\
9&0.85&0.66&4.14&0.55&&1.35&0.86&3.56&0.84&&1.35&0.94&3.59&0.90\\
H&1.41&1.16&5.57&0.72&&2.67&2.07&6.27&1.15&&2.65&2.15&6.24&1.19\\
H-L&1.94&1.56&3.10&1.74&&4.15&3.03&4.45&2.35&&3.99&3.11&3.98&2.71\\
\toprule
&\multicolumn{4}{c}{NN3}&&\multicolumn{4}{c}{NN4}&&\multicolumn{4}{c}{NN5}\\
Decile&Pred&Real&Std&SR &&Pred&Real&Std&SR &&Pred&Real&Std&SR\\
\midrule
L&-1.28&-0.67&6.10&-0.38&&-1.22&-0.68&6.12&-0.39&&-0.80&-0.55&5.84&-0.33\\
2&-0.42&0.07&4.68&0.05&&-0.30&0.04&4.63&0.03&&-0.07&0.12&4.45&0.10\\
3&-0.05&0.25&3.79&0.23&&0.09&0.26&3.79&0.24&&0.20&0.35&3.75&0.33\\
4&0.20&0.30&3.48&0.30&&0.34&0.34&3.40&0.34&&0.39&0.42&3.33&0.44\\
5&0.40&0.40&3.31&0.42&&0.54&0.42&3.23&0.45&&0.54&0.51&3.28&0.54\\
6&0.59&0.53&3.21&0.58&&0.73&0.54&3.18&0.58&&0.68&0.46&3.21&0.50\\
7&0.78&0.62&3.21&0.67&&0.93&0.64&3.19&0.69&&0.83&0.60&3.21&0.65\\
8&1.00&0.76&3.25&0.81&&1.16&0.74&3.32&0.78&&1.01&0.67&3.30&0.70\\
9&1.32&0.91&3.51&0.90&&1.48&0.87&3.56&0.84&&1.28&0.76&3.74&0.70\\
H&2.66&2.04&6.04&1.17&&2.92&2.05&6.02&1.18&&2.65&1.86&6.40&1.01\\
H-L&4.20&2.98&4.31&2.39&&4.40&3.00&4.13&2.51&&3.71&2.68&4.27&2.17\\
\bottomrule
\end{tabular}
}
\caption{Performance of Value Weighted Portfolios by Decile.}
\label{tab:decile_value}
\end{center}
\vspace{-0.2cm}
Note: \textit{Performance on decile of each model for value weighted portfolios. Rows L, H and H-L stand for Low, High and High-minus-Low deciles respectively. Columns Pred, Real, Std, and SR are average predicted monthly returns (in \%), average realised monthly returns (in \%), realised monthly standard deviation (in \%) and Sharpe ratio, respectively.}
\end{table}

\begin{table}[]
    \begin{center}
    \resizebox{\textwidth}{!}{
    \begin{tabular}{l rrrr rrrr rrrr r} %cccc cccc cccc c }
    \toprule
    &OLS&OLS3&PLS&PCR&ENet&GLM&RF&GBRT&NN1&NN2&NN3&NN4&NN5\\
    &+H&+H&&&+H&+H&&+H&&&&&\\
    \midrule
    &\multicolumn{13}{c}{Equally Weighted}\\
    Ann Ret&0.36&0.21&0.31&0.30&0.30&0.32&0.28&0.25&0.48&\textbf{0.50}&0.48&0.48&0.43\\
    % Cum. Ret.&10.76&6.32&9.24&9.01&9.09&9.57&8.31&7.60&14.54&\textbf{15.11}&14.43&14.51&13.02\\
    Vol&0.16&0.19&0.17&0.17&0.17&0.18&\textbf{0.14}&0.15&0.20&0.18&0.20&0.19&0.20\\
    SR&2.28&1.11&1.85&1.78&1.77&1.81&1.96&1.71&2.39&\textbf{2.74}&2.42&2.56&2.21\\
    Skew&0.52&0.77&0.14&0.46&-0.05&-0.26&0.97&1.72&2.18&2.27&1.69&1.75&\textbf{2.40}\\
    Kurt&\textbf{4.36}&17.89&7.96&7.50&7.42&8.63&7.03&12.63&19.44&13.94&11.12&11.47&19.56\\
    Max DD&0.25&0.63&0.44&0.31&0.40&0.40&0.23&0.24&0.28&\textbf{0.17}&0.29&0.23&0.47\\
    Max Loss&0.13&0.36&0.23&0.20&0.22&0.27&0.17&0.16&\textbf{0.28}&0.16&0.23&0.22&0.23\\
    Turnover&1.26&1.50&1.15&1.27&1.28&1.36&\textbf{0.92}&1.25&1.24&1.23&1.20&1.20&1.15\\
    &\multicolumn{13}{c}{Value Weighted}\\
    Ann Ret&0.27&0.16&0.23&0.23&0.23&0.24&0.21&0.19&0.36&\textbf{0.37}&0.36&0.36&0.32\\
    Vol&0.12&0.14&0.12&0.12&0.13&0.13&\textbf{0.10}&0.11&0.16&0.14&0.15&0.14&0.15\\
    SR&2.29&1.13&1.84&1.83&1.79&1.84&2.04&1.72&2.33&\textbf{2.67}&2.37&2.48&2.16\\
    Skew&0.84&0.82&0.16&0.62&0.13&-0.21&1.28&1.89&2.39&2.48&1.87&1.90&\textbf{2.71}\\
    Kurt&\textbf{5.30}&23.68&9.14&7.84&7.58&9.78&6.52&14.23&23.00&16.41&13.35&13.26&24.58\\
    Max DD&0.16&0.51&0.36&0.21&0.28&0.33&0.16&0.20&0.23&\textbf{0.14}&0.24&0.19&0.38\\
    Max Loss&\textbf{0.09}&0.30&0.19&0.14&0.16&0.22&0.10&0.12&0.23&0.13&0.19&0.18&0.19\\
    Turnover&1.05&1.33&1.02&1.06&0.97&1.07&\textbf{0.51}&0.67&0.66&0.60&0.69&0.55&0.52\\
    \bottomrule
    \end{tabular}
    }
    \end{center}
    \caption{Statistical Performance of Expert Portfolios.}
Note: \textit{Statistical performance of equally and value weighted portfolios. Columns Ann Ret, Vol, SR, Skew, Kurt,  Max DD, Max Loss and Turnover stand for annualised average return, annualised volatility, annualised Sharpe ratio, skewness, kurtosis, maximum drawdown,  1-month maximum loss and portfolio turnover. The metrics are computed on the test period 1987-2016. }
\label{tab:ptfperf_value}
\end{table}

\begin{table}[]
    \begin{center}
    \resizebox{\textwidth}{!}{
    \begin{tabular}{l ccc ccc}
    \toprule
    &Best Expert&\multicolumn{2}{c}{Fixed Combination}&\multicolumn{2}{c}{Adaptative Mixture}& Oracle\\
    \cmidrule(lr){3-4} \cmidrule(lr){5-6}
    &NN2&PtfUNI&Best Convex&Best Convex&PtfBOA&Best Convex\\
    &&&on Valid. Set&One-year Rolling&&\\
    \midrule
    &\multicolumn{6}{c}{Equally Weighted}\\
    Ann. Ret.&\textbf{0.50}&0.36&0.36&0.43&0.49&0.50\\
    % Cum. Ret.&\textbf{14.36}&10.45&10.25&12.24&14.03&14.38\\
    Vol.&0.18&\textbf{0.14}&0.16&0.16&0.18&0.17\\
    SR&2.74&2.56&2.28&2.60&\textbf{2.77}&2.92\\
    Skew.&2.27&1.19&0.52&1.65&\textbf{3.11}&2.90\\
    Kurt.&13.94&10.17&\textbf{4.36}&10.58&19.63&17.93\\
    Max DD&0.17&0.24&0.25&0.17&\textbf{0.08}&0.07\\
    Max Loss&0.16&0.18&0.13&0.16&\textbf{0.08}&0.07\\
    Turnover&1.23&\textbf{1.22}&1.26&\textbf{1.22}&1.23&1.24\\
    &\multicolumn{6}{c}{Value Weighted}\\
    Ann. Ret.&\textbf{0.37}&0.27&0.27&0.32&0.36&0.37\\
    % Cum. Ret.&10.77&7.89&7.78&9.17&10.36&10.75\\
    Vol.&0.14&\textbf{0.11}&0.12&0.12&0.13&0.13\\
    SR&2.67&2.54&2.29&2.57&\textbf{2.73}&2.76\\
    Skew&2.48&1.20&0.84&1.69&\textbf{3.00}&3.21\\
    Kurt&16.41&10.73&\textbf{5.30}&8.95&18.85&22.20\\
    Max DD&0.14&0.20&0.16&0.14&\textbf{0.09}&0.08\\
    Max Loss&0.13&0.15&0.09&0.11&\textbf{0.05}&0.08\\
    Turnover&\textbf{0.52}&0.82&1.05&0.75&0.79&0.73\\
    \bottomrule
    \end{tabular}
    }
    
    \end{center}
    \caption{Performance of Aggregated Portfolios.}
Note: \textit{Columns Ann. Ret., Vol., Skew., Kurt., SR, Max DD, Max Loss and Turnover stand for annualised average return, volatility, skewness, kurtosis, annual Sharpe ratio, maximum drawdown,  1-month maximum loss and portfolio turnover. The metrics are computed on the test period 1987-2016. 
Expect for NN2, all the portfolios are convex combinations of the thirteen experts. The oracle is the best possible convex mixture on the test period and is unachievable in practice. \\ }
\label{tab:stratstatdesc_value}
\end{table}

\begin{table}[]
    \begin{center}
    \includegraphics[width=\textwidth]{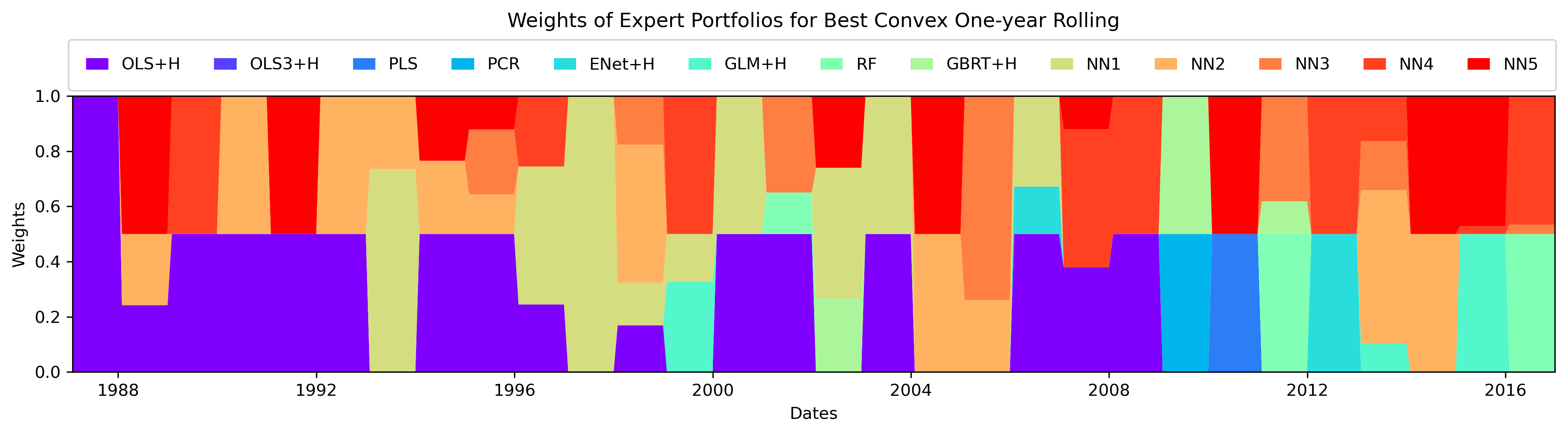}
    \resizebox{\textwidth}{!}{
    \begin{tabular}{l cccc cccc cccc c}
    \toprule
    &OLS&OLS3&PLS&PCR&ENet&GLM&RF&GBRT&NN1&NN2&NN3&NN4&NN5\\
    &+H&+H&&&+H&+H&&+H&&&&&\\
    \midrule
    &\multicolumn{13}{c}{Equally Weighted}\\
    PtfUNI&0.08&0.08&0.08&0.08&0.08&0.08&0.08&0.08&0.08&0.08&0.08&0.08&0.08\\
    Best Convex on Valid. Set&1.00&0.00&0.00&0.00&0.00&0.00&0.00&0.00&0.00&0.00&0.00&0.00&0.00\\
    Best Convex One-year Rolling&0.25&0.00&0.02&0.02&0.02&0.03&0.04&0.03&0.15&0.14&0.07&0.11&0.12\\
    PtfBOA&0.23&0.01&0.00&0.00&0.01&0.02&0.00&0.00&0.09&0.44&0.08&0.07&0.03\\
    Oracle&0.30&0.00&0.00&0.00&0.00&0.00&0.00&0.00&0.00&0.59&0.00&0.11&0.00\\
    &\multicolumn{13}{c}{Value Weighted}\\
    PtfUNI&0.08&0.08&0.08&0.08&0.08&0.08&0.08&0.08&0.08&0.08&0.08&0.08&0.08\\
    Best Convex on Valid. Set&1.00&0.00&0.00&0.00&0.00&0.00&0.00&0.00&0.00&0.00&0.00&0.00&0.00\\
    Best Convex One-Year Rolling &0.25&0.01&0.02&0.01&0.03&0.03&0.03&0.03&0.15&0.13&0.07&0.11&0.13\\
    PtfBOA&0.38&0.01&0.00&0.00&0.01&0.01&0.00&0.00&0.06&0.39&0.05&0.06&0.02\\
    Oracle&0.26&0.00&0.00&0.00&0.00&0.00&0.00&0.00&0.24&0.50&0.00&0.00&0.00\\

    \bottomrule
    \end{tabular}
    }
    \end{center}
    \caption{Average Weights of the Mixtures.}
    \label{tab:weights}
Note: \textit{
Average weights of the mixture portfolios on the test period 1987-2016. 
PtfUNI, Best Convex on Valid. Set, Best Convex One-year Rolling, PtfBOA, Oracle indicate the uniform mixture, the best fixed convex combination on the last year of the validation set, the one-year rolling fixed combination, BOA and the oracle portfolio weights respectively.
The rolling best convex and BOA are adaptive mixtures, the weight evolution of the rolling mixture is given in the graph.
Oracle is the best possible convex mixture on the test period and is unachievable in practice.
}
\end{table}

\begin{table}[H]
\begin{center}
\resizebox{.8\textwidth}{!}{
\begin{tabular}{l ccc ccc c}
\toprule
Experts&Ann Ret&Vol&Skew&Kurt&SR&Max DD& Max Loss \\
\midrule
NN2\_0&0.37&0.16&1.27&7.46&2.27&0.19&0.18\\
NN2\_1&0.40&0.18&1.00&7.31&2.27&0.25&0.25\\
NN2\_2&0.38&0.17&1.06&7.34&2.25&0.35&0.22\\
NN2\_3&0.39&0.20&0.77&8.97&1.91&0.42&0.26\\
NN2\_4&0.37&0.20&0.55&10.19&1.89&0.83&0.29\\
NN2\_5&0.37&0.17&1.41&7.89&2.11&0.19&0.10\\
NN2\_6&0.40&0.20&1.15&8.89&1.97&0.51&0.22\\
NN2\_7&0.36&0.18&2.83&22.44&2.06&0.23&0.14\\
NN2\_8&0.39&0.19&2.39&22.49&2.00&0.76&0.26\\
NN2\_9&0.39&0.17&1.22&10.61&2.26&0.41&0.22\\
OLS+H\_0&0.33&0.15&0.12&5.05&2.17&0.33&0.16\\
OLS+H\_1&0.32&0.16&0.22&6.54&1.97&0.24&0.23\\
OLS+H\_2&0.30&0.16&-0.05&5.20&1.92&0.41&0.19\\
OLS+H\_3&0.32&0.15&-0.30&7.40&2.16&0.26&0.26\\
OLS+H\_4&0.32&0.15&-1.03&10.1&2.10&0.31&0.29\\
OLS+H\_5&0.31&0.14&0.24&3.32&2.13&0.21&0.14\\
OLS+H\_6&0.35&0.16&0.53&3.72&2.23&0.22&0.14\\
OLS+H\_7&0.32&0.15&0.60&5.28&2.17&0.26&0.13\\
OLS+H\_8&0.32&0.15&0.30&5.52&2.11&0.36&0.18\\
OLS+H\_9&0.29&0.18&0.19&5.75&1.59&0.41&0.19\\
% &&&&&&\\
Extended UNI&0.36&0.13&1.13&8.15&2.79&0.19&0.12\\
Extended BOA&0.47&0.17&2.98&18.39&2.82&0.09&0.07\\
\bottomrule
\end{tabular}
}
\end{center}
\caption{Performance of Specialised Expert Portfolios and Extended Mixtures.}
Note: \textit{Columns Ann Ret, Vol, Skew, Kurt, SR, Max DD, Max Loss provide annualised mean return, volatility, Skewness, Kurtosis, annual sharpe ratio, maximum drawdown and 1-month maximum loss. Extended BOA includes the $K=13$ initial experts plus $K^{'}=10$ new specialised neural networks NN2 and $K^{'}=10$ new specialised OLS+H. 
% The performances of the initial set of experts stay the same as in Table \ref{tab:stratstatdesc}. 
Portfolios are equally weighted.
}
\label{tab:spedesc}
\end{table}

\end{document}